\documentclass{aa}  

\usepackage{graphicx}
\usepackage{txfonts}
\begin{document}
   \title{Monitoring the temperature and reverberation delay of the circumnuclear hot dust in NGC 4151}


   \author{K.~Schn\"ulle\inst{1},
	  J.-U.~Pott\inst{1},
	  H.-W.~Rix\inst{1},
	  B.~M.~Peterson\inst{2,3},
	  G.~De Rosa\inst{2,3,4}	  
          \and
          B.~Shappee\inst{2,5}
          }
	  
   \institute{Max Planck Institut f\"ur Astronomie, K\"onigstuhl 17, D-69117
Heidelberg, Germany,
              \email{schnuelle@mpia.de}
\and
	      Department of Astronomy, The Ohio State University, 140 W 18th
Ave, Columbus, OH 43210, USA
\and
  Center for Cosmology and AstroParticle Physics, The Ohio State University, 191 
West Woodruff Avenue, Columbus, OH 43210, USA
\and
  Space Telescope Science Institute, 3700 San Martin Drive, Baltimore, MD 21218, USA
\and
Carnegie Observatories, 813 Santa Barbara Street, Pasadena, CA 91101, USA}

   \date{\today}

 
  \abstract
   {A hot, dusty torus located around the outer edge of the broad-line
region of AGNs is a fundamental ingredient in unified AGN models.
While the existence of circumnuclear dust around AGNs at pc-scale radii is now widely accepted, questions about the origin, evolution and long-term stability of these dust tori remain unsettled.\\ 
We used reverberation mapping of the hot circumnuclear dust in the Seyfert 1 galaxy NGC 4151, to monitor its temperature and reverberation lag as a function of the varying accretion disk brightness.
We carried out multiband, multiepoch photometric observations of the nucleus of NGC 4151 in the
$z,Y,J,H,$ and $K$ bands for 29 epochs from 2010 January to 2014 June, supported by new near-infrared and optical spectroscopic observations, and archived WISE data.\\
We see no signatures of dust destruction due to sublimation in our data, since they show no increase in the hot dust reverberation delay directly correlated with substantial accretion disk flux increases in the observed period.
Instead, we find that the hot dust in NGC 4151 appears to merely heat up, and the hot dust temperature closely tracks the accretion disk luminosity variations. We find indications of a decreased reverberation delay within the observed period from $\tau=42.5 \pm 4.0$ days in 2010 to $\tau=29.6 \pm 1.7$ days in 2013-2014.
Such a varying reverberation radius on longer timescales would explain the intrinsic scatter observed in the radius-luminosity relation of dust around AGNs.\\
Our observations rule out that a second, larger dust component within a 100-light-day radius from the source contributes significantly to the observed near-infrared flux in this galaxy.}

 {}
   {}
   {}
   {}

\titlerunning{Dust reverberation in NGC 4151}
\authorrunning{Schn\"ulle et al.}

   \maketitle
%


\section{Introduction}
\label{sec:Introduction}

To explain the apparent dichotomy between Type 1 and Type 2 AGNs, the unified AGN model postulates a so-called dust
torus \citep{Ant1993,Urr1995} around the broad-line region (BLR) of AGNs. Depending on viewing angle, this optically thick structure
will, or will not, obscure the central region, resulting
in different spectral energy distributions (SEDs) for Type 1 and Type 2 AGNs.
Direct and indirect
observational evidence \citep{Bar1987,Hei1997,Ost2006,Jaf2004,Eli2006} has been accumulated over past decades, confirming the existence of the assumed dusty torus around the BLR.
Nevertheless, our knowledge of the long-term stability of the dust around AGNs and of its origin remains vague.\\
Self-consistent AGN torus models \citep{Kro1988,Scha2010} assume that dust is
ejected into the
interstellar medium by asymptotic giant branch (AGB) stars and 
brought from
outside to the
central nuclear region.
Unless the luminosity of the central source has decreased on a timescale shorter than the dynamical timescale, the innermost dust for these models is always expected at the sublimation radius.
A different scenario, which suggests a
location or formation
of dust farther out than the sublimation radius, has been proposed by
\citet{Elv2002}. Here, the dust formation results from BLR clouds
that are assumed to be part of an outflowing accretion disk (AD) wind\footnote{It is not yet
understood well \citep{Pet2006b} whether the BLR clouds are infalling,
outflowing
or in rotation.
Recent observations indicate that
the signatures of low-ionization broad emission lines are consistent with gas
that is infalling or in a rotating Keplerian disk \citep{Gri2013,Pet2006b},
whereas the higher ionized lines seem to trace gas that is part of an
outflowing AD wind
\citep{Cren1999,Pet2006a,Kol2013}.}. 
In
the course
of a subsequent expansion and cooling process, the conditions of dust
condensation are
fulfilled and dust is formed -- by the AGN itself -- at a distance farther out
than the sublimation radius.\\ 
To be able to distinguish between competing dust
formation
scenarios, it seems essential to determine the stability and evolution of the innermost dust radius. Monitoring the hot dust temperature is a further indicator of the immediate dust state, i.e.~whether the hot dust is at the sublimation temperature.
Observations, even if performed on one and the same AGN, NGC 4151, have so far given seemingly inconclusive indications:
\begin{itemize}
\item $V-K$
reverberation measurements indicate strongly varying time lags 
between 30 and 70 days over the years 2001 - 2006 \citep{Kosh2009},
 suggesting 
a strong variation in the location of the innermost 
dust over the observed period, possibly by means of dust destruction and fast re-formation.
A reanalysis of these data by \citet{Hoe2011} with a simplified
clumpy torus model resulted in an apparently constant time lag of $\tau = 43.8 \pm 8.5$ days for that same period. 
\item Interferometric $K$-band observations in 2010 \citep{Pot2010} have shown that
the hot circumnuclear dust torus around NGC 4151 apparently does not expand
owing to dust sublimation when the AD brightens, but found indications of dust survival in bright times.
\item From our 2010 $z$ to $K$ photometry of NGC 4151, we
found no signatures of dust destruction following increased AD emission (\citealt{Schn2013}, hereafter Paper I). Instead, we measured a significant increase in the hot dust
temperature, indicating that the 
hot dust in NGC 4151 was apparently outside the current sublimation radius in that period. 
\item Recently, with $K$-band interferometric observations (Oct
2010 - May
2012), \citet{Kish2013}
have found indications of an expanding $K$-band dust location for
NGC4151 caused by increased
AD emission and a correlation of the hot dust radius with the averaged AD flux over the past six years.
\end{itemize}  
While generally the dust
location around AGNs is
known to correlate with the average AD luminosity as $R_{dust}\propto\sqrt{L_{AD}}$ \citep{Sug2006,Kish2011a}, 
the above results imply that it is not a tight
function of the instantaneous luminosity state. \\
We present here the updated results from our monitoring program of NGC 4151 during 2010-2014, in which we used dust reverberation to study the evolution of the temperature and reverberation lag of the hot dust
around this archetypical Sy1 AGN. Compared to Paper I, we substantially extended the data set by using a significantly longer time series and a broader wavelength coverage. We also extended our reduction and analysis techniques by performing precise photometry with an image subtraction method and employing a multiepoch, multiwavelength Markov chain Monte Carlo (MCMC) fit. NGC 4151 is part of our AGN
hot dust reverberation campaign, in which we monitor roughly 25 nearby, bright, and
variable Type 1 AGNs from the optical to the near-infrared (NIR). \\
Observations, data reduction and photometry are described in
Sect.~\ref{sec:Observations}, followed by an introduction of our data analysis methods in Sect.~\ref{sec:Methods}. We present our updated results on NGC 4151 in Sect.~\ref{sec:Results}, and discuss these in Sect.~\ref{sec:Discussion}. A summary of our results, as well as the outlook for this project, is given in Sect.~\ref{sec:Conclusions}.

\section{Observations and data processing} 
\label{sec:Observations}
\subsection{Near-infrared data}

\begin{table*}
\caption{NIR fluxes of the nucleus of NGC 4151, derived with {\tt ISIS}.}       
\label{tab:ISIS}      
\centering          
\begin{tabular}{ c | c | c | c | c | c | c }     
\hline       
Epoch & MJD & \multicolumn{5}{c}{$F_\lambda$ / $10^{-13} W\mbox{m}^{-2}\mu\mbox{m}^{-1}$ } \\ 
 & & z & Y & J & H & K\\ 
\hline                 
 1 & 55227   &  1.321   $\pm$   0.103   &   1.107   $\pm$   0.074   &   1.075   $\pm$   0.039   &   0.970    $\pm$  0.032   &   0.965  $\pm$    0.034\\
 2 & 55252   &   1.577   $\pm$  0.068   &   1.324   $\pm$  0.052    &  1.284    $\pm$  0.043    &  1.239   $\pm$  0.045    & 1.127    $\pm$  0.051\\
 3 & 55283   &   1.691   $\pm$   0.070  &    1.469   $\pm$   0.050  &    1.489  $\pm$    0.051  &    1.499   $\pm$   0.054  &    1.397   $\pm$   0.034\\
 4 & 55311   &   1.733   $\pm$  0.059   &   1.550   $\pm$   0.044   &   1.585   $\pm$   0.071   &   1.637   $\pm$   0.052   &   1.529   $\pm$   0.080\\
 5 & 55344   &   1.573   $\pm$   0.085  &   1.431  $\pm$    0.058   &  1.536   $\pm$  0.043     & 1.723   $\pm$   0.049     & 1.643  $\pm$   0.056\\
 6 & 55373   &   1.530   $\pm$  0.083   &   1.390   $\pm$   0.042   &   1.424  $\pm$    0.064   &   1.546  $\pm$    0.050   &   1.543   $\pm$   0.056\\
 7 & 55969   &  0.739    $\pm$  0.075   &  0.516   $\pm$  0.056     & 0.625   $\pm$   0.060     & 0.712    $\pm$  0.083     & 0.802   $\pm$   0.037\\
 8 & 55998   &   0.630   $\pm$   0.050  &    0.483   $\pm$ 0.037    &  0.539   $\pm$   0.056    &  0.605   $\pm$   0.040    &  0.642    $\pm$  0.041\\
 9 & 56024   &   0.950   $\pm$   0.032  &    0.749   $\pm$   0.039  & 0.714   $\pm$   0.064 &  0.653  $\pm$    0.073  &    0.670  $\pm$    0.042\\
 10 & 56255   &   1.270   $\pm$   0.061  &    1.038  $\pm$    0.042 &     1.080    $\pm$  0.060 &     1.068   $\pm$   0.040  &   1.004   $\pm$   0.033\\
 11 & 56264   &  1.030    $\pm$  0.034   &   0.881   $\pm$   0.033    &  0.971  $\pm$    0.031   &   0.973   $\pm$   0.030    &  0.969   $\pm$  0.050\\
 12 & 56285   &   1.193   $\pm$   0.082  &   1.195  $\pm$    0.172    & 1.089   $\pm$  0.060     & 1.116   $\pm$   0.035     & 1.086   $\pm$   0.039\\
 13 & 56319   &   1.169   $\pm$   0.040  &    1.047   $\pm$   0.035   &   1.103   $\pm$   0.033  &    1.161   $\pm$   0.039   &   1.199   $\pm$   0.037\\
 14 & 56374   &   0.953   $\pm$   0.074  &    0.770   $\pm$   0.045   &   0.826   $\pm$   0.064  &    0.889   $\pm$   0.064   &   0.915 $\pm$     0.033\\
 15 & 56436   &   1.007   $\pm$   0.030  &    0.830   $\pm$   0.030   &   0.901  $\pm$   0.030   &   0.977   $\pm$   0.030    &  0.966   $\pm$   0.030\\
 16 & 56465   &   0.721   $\pm$   0.037  &    0.570  $\pm$    0.048  &    0.746   $\pm$   0.039 &     0.827  $\pm$    0.032  &    0.913    $\pm$  0.039\\
 17 & 56500    &  0.558   $\pm$   0.035  &    0.389  $\pm$   0.055    &  0.526  $\pm$   0.043    &  0.626   $\pm$   0.059     & 0.701    $\pm$  0.039\\
 18 & 56523    &  0.613   $\pm$   0.067  &   0.462   $\pm$   0.063    &  0.578   $\pm$   0.057   &   0.570   $\pm$   0.048    &  0.648   $\pm$   0.038\\
 19 & 56586    &  0.914   $\pm$   0.039  &    0.732   $\pm$   0.035   &   0.791  $\pm$    0.079  &    0.781   $\pm$   0.080   &   0.857  $\pm$    0.048\\
 20 & 56611    &  0.917    $\pm$  0.031  &    0.736   $\pm$   0.058  &    0.862   $\pm$   0.032 &     0.916  $\pm$   0.033   &   0.909  $\pm$   0.071\\
 21 & 56617    &  0.810    $\pm$  0.034  &    0.675  $\pm$    0.074   &   0.792  $\pm$    0.031  &    0.885  $\pm$   0.032    & 0.921  $\pm$    0.040\\
 22 & 56669    &  0.500  $\pm$    0.036  &   0.393   $\pm$   0.037    &  0.453  $\pm$   0.055    &  0.592   $\pm$  0.035    & 0.729   $\pm$  0.045\\
 23 & 56732    &  0.498   $\pm$   0.039  &   0.393   $\pm$   0.042    &  0.378   $\pm$   0.034   &   0.358   $\pm$   0.044  &    0.413   $\pm$   0.051\\
 24 & 56737    &  0.559   $\pm$   0.034  &   0.392   $\pm$  0.040   &  0.400  $\pm$    0.039    & 0.357   $\pm$   0.048     & 0.413   $\pm$   0.044\\
 25 & 56763    &  0.569   $\pm$  0.040   &   0.394   $\pm$   0.051  &    0.453   $\pm$   0.030  &    0.434  $\pm$    0.038  &   0.469   $\pm$   0.034\\
 26 & 56787    &  0.505   $\pm$  0.056   &   0.418  $\pm$   0.031   &   0.453  $\pm$    0.040   &   0.482   $\pm$   0.042   &  0.520 $\pm$    0.037\\
 27 & 56792    &  0.501   $\pm$   0.032 &     0.380   $\pm$   0.038 &    0.453  $\pm$    0.035 &     0.465  $\pm$    0.044  &    0.533  $\pm$    0.042\\
 28 & 56797    &  0.560   $\pm$   0.043  &    0.392   $\pm$   0.055  &    0.464   $\pm$   0.043 &     0.482  $\pm$    0.043  &    0.526  $\pm$    0.042\\
 29 & 56826    &  0.551   $\pm$   0.049  &    0.429  $\pm$    0.043  &    0.478  $\pm$    0.047 &     0.482   $\pm$   0.030  &    0.532  $\pm$   0.030\\

\hline              
\end{tabular}
\end{table*}

\begin{figure}
\includegraphics[width=9.cm,angle=0,clip=true]{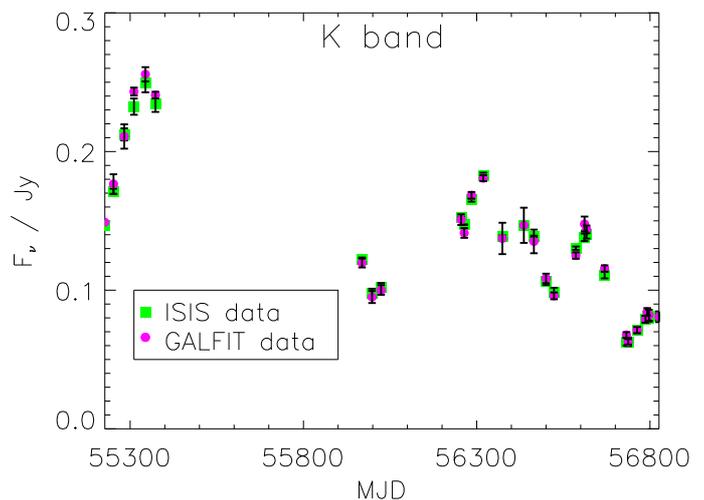}
\caption{\label{fig:ISISvsGalfit} 
Nuclear fluxes of NGC 4151 derived with {\tt ISIS} (green) and {\tt GALFIT} (magenta), shown here for the $K$ band. Within the photometric errors, we observe excellent agreement between the two methods. The agreement is on a similar level for all other bands, suggesting that both methods manage to separate the nuclear flux of interest from the bulge and disk.} 
\end{figure}

From 2010 January - June, we obtained NIR multiband photometric observations of the Sy1 NGC 4151 (see Paper I), using the Omega 2000 \citep{Kov2004} NIR
wide-field
camera  mounted on the 3.5m telescope in Calar Alto,
Spain\footnote{Based on
observations collected at the Centro Astronómico Hispano Alemán (CAHA), operated
jointly
by the Max-Planck Institut für Astronomie and the Instituto de Astrofisica de
Andalucia
(CSIC).}. 
Omega 2000 has a field of view $15\farcm4 \times 15\farcm4$, and its 2048 $\times$ 2048
pixel detector has a scale of $0\farcs45$ pixel$^{-1}$.\\ 
We continued our monitoring of NGC 4151 from 
2012 February - 2014 June, with roughly two to four weeks of sampling (apart from data gaps in 2010 July - 2012 February and 2012 May - November). In each of the 29 epochs,
we obtained broadband photometry of NGC 4151 and of three calibration stars in the same
field of view in the
$z$, $Y$, $J$, $H$, and $K$ spectral bands.
These five filters were chosen to differentiate between variations in the AD and hot dust emission. 
The weather conditions ranged from clear to photometric,
and the seeing was $0\farcs9-2\farcs5$.\\
Following the data reduction with IRAF (see Paper I), we extracted the nuclear fluxes of NGC
4151 at various epochs and filters, using two different approaches. First, we performed a PSF-bulge-disk decomposition with {\tt GALFIT}
\citep{Pen2002},
which allows for the
simultaneous fitting of different galaxy components to the objects in an
image.
For NGC 4151, the fit model consisted of the point spread function (PSF) --
constructed with the help of the {\tt IRAF
DAOPHOT} package for each frame -- plus a Sersic profile and an exponential disk profile. This
strategy led to
 small
statistical errors in the fit (typically 0.01 mag for the
PSF and Sersic components, and 0.02 mag for
the disk component)
as well as sufficiently smooth residuals for our purpose.
The Sersic half-light radii $R_e$, Sersic indices $n$, and disk
scale lengths $h$ for the different filters and epochs were found to be temporally stable. In a final fit, $R_e$, $n$, and $h$ were fixed to their average values determined from the previous fits, i.e.~$R_e=9\farcs0$, $n=3.0$, $h=41\farcs8$.
These parameter values for NGC 4151 are consistent with the literature, for example \citet{Ben2009}.  
The residuals deviate significantly from zero only in the PSF regime, while
they are very smooth on larger scales. Substantial residuals in the PSF domain
are likely to be caused by the complex PSF (due to defocusing to avoid saturation,
plus seeing), which is variable over the detector. 
Aperture photometry on these residuals yielded low fluxes, on the
order
of 2-3 \% of the PSF flux of the AGN.
The fluxes of the reference
stars were determined by performing a simultaneous PSF model fit in {\tt GALFIT}. \\
Absolute flux was calibrated in the $JHK$ bands with the known fluxes of our calibrators, as derived from the 2MASS PSC
\citep{Skru2006}. For the $zY$
photometry, 
we used a database
containing model spectra of main sequence stars (kindly provided by R.~van
Boekel, priv.~com.). For each calibrator, we determined the best fit spectrum to the $BVJHK$ fluxes as given by
SIMBAD and calculated the $zY$ photometry from that spectrum.\\
As an alternative to {\tt GALFIT}, we used the {\tt ISIS} image subtraction package \citep{Alard1998,Alard2000}, which enables precise measurement of flux variations in variable objects. Following the procedures described by
\cite{Shappee2011}, the images in each band are first aligned using the program {\tt
Sexterp} \citep{Siverd2012}. 
Then a reference image has to be chosen or constructed -- in our case, we chose the image with the best seeing. A spatially variable convolution kernel $K$ is determined that minimizes the discrepancy
\begin{equation}
D=\sum_i ([R \otimes K](x_i,y_i)-I(x_i,y_i))^2\, ,
\end{equation}
i.e.~the kernel optimally transforms the reference image $R$ to the seeing and flux level of a given individual image $I$.
Each frame is then subtracted from the convolved
reference image. Light curves, which are difference fluxes relative to the reference image, are then extracted from the
subtracted images by performing aperture photometry on the
residual nuclear flux. The error in the difference flux is estimated by measuring the residual flux of non-variable stars in the field (which should ideally be zero). Absolute calibration of the difference flux of each individual frame was performed with the same three calibration stars as in our {\tt GALFIT} analysis. The absolute nuclear flux level in the reference image was also adopted from our {\tt GALFIT} results. As can be seen in Fig.~\ref{fig:ISISvsGalfit} for the $K$ band, our {\tt GALFIT} and {\tt ISIS} fluxes agree well within the given errors. For targets at substantially farther distances than NGC 4151 (as are present in our broad AGN sample), a {\tt GALFIT} decomposition might be more difficult owing to increased degeneracies between PSF and the bulge, making {\tt ISIS} the method of choice for these objects. For this reason and overall consistency, we use the {\tt ISIS} photometry (listed in Table \ref{tab:ISIS}) throughout this paper.\\
To remove emission line contributions in our $zYJHK$ fluxes, a
NIR spectrum of NGC 4151  was obtained with the SpeX spectrograph
\citep{Ray2003} at the NASA 
Infrared Telescope Facility (IRTF) on 2010 February 26, as already described in Paper I. For each
filter, we applied a correction
factor to our data ($\approx$8\% in $zY$, $\approx$15\% in $J$,
 $\approx$2\% in $HK$), derived from the ratio
of the synthetic IRTF
photometry to the
continuum flux level at the respective effective wavelength.\\

\subsection{Mid-infrared data}
\begin{table*}
\caption{WISE W1, W2, W3 photometric fluxes of NGC 4151, derived with {\tt GALFIT}. We find very good agreement with the total NGC 4151 fluxes
as listed in the ALLWISE catalog (column 5). Our total fluxes are systematically slightly lower than the catalog fluxes, which is partly caused by the correction term that we applied to shift our fluxes from the mean epoch to 2010 May (see text for details). From 2010 May-December, the AGN got approximately 25\% brighter.}       
\label{tab:WISE}      
\centering          
\begin{tabular}{ l | c | c | c | c }     
\hline       
\hspace{-0.15cm}Filter &  \multicolumn{4}{c}{$F_{\lambda}$ / $10^{-13} W\mbox{m}^{-2}\mu\mbox{m}^{-1}$}\\
\hspace{-0.15cm} & AGN & host
& total & total acc.~ALLWISE\\  
\hline                 
\hspace{-0.15cm}W1 &  0.685 $\pm$ 0.094   &   0.625 $\pm$ 0.077   &	1.311 $\pm$ 0.171   & 1.625 $\pm$ 0.088\\ 
\hspace{-0.15cm}W2 &  0.538 $\pm$ 0.068  &   0.301 $\pm$ 0.101  &	0.839 $\pm$ 0.170   & 1.022 $\pm$ 0.044\\ 
\hspace{-0.15cm}W3 & 0.192 $\pm$ 0.035  &    0.076 $\pm$ 0.011   &	0.268 $\pm$ 0.047  &	0.311 $\pm$ 0.028 \\

\hline              
\end{tabular}
\end{table*}

\begin{figure*}
\includegraphics[width=18.cm,angle=0,clip=true]{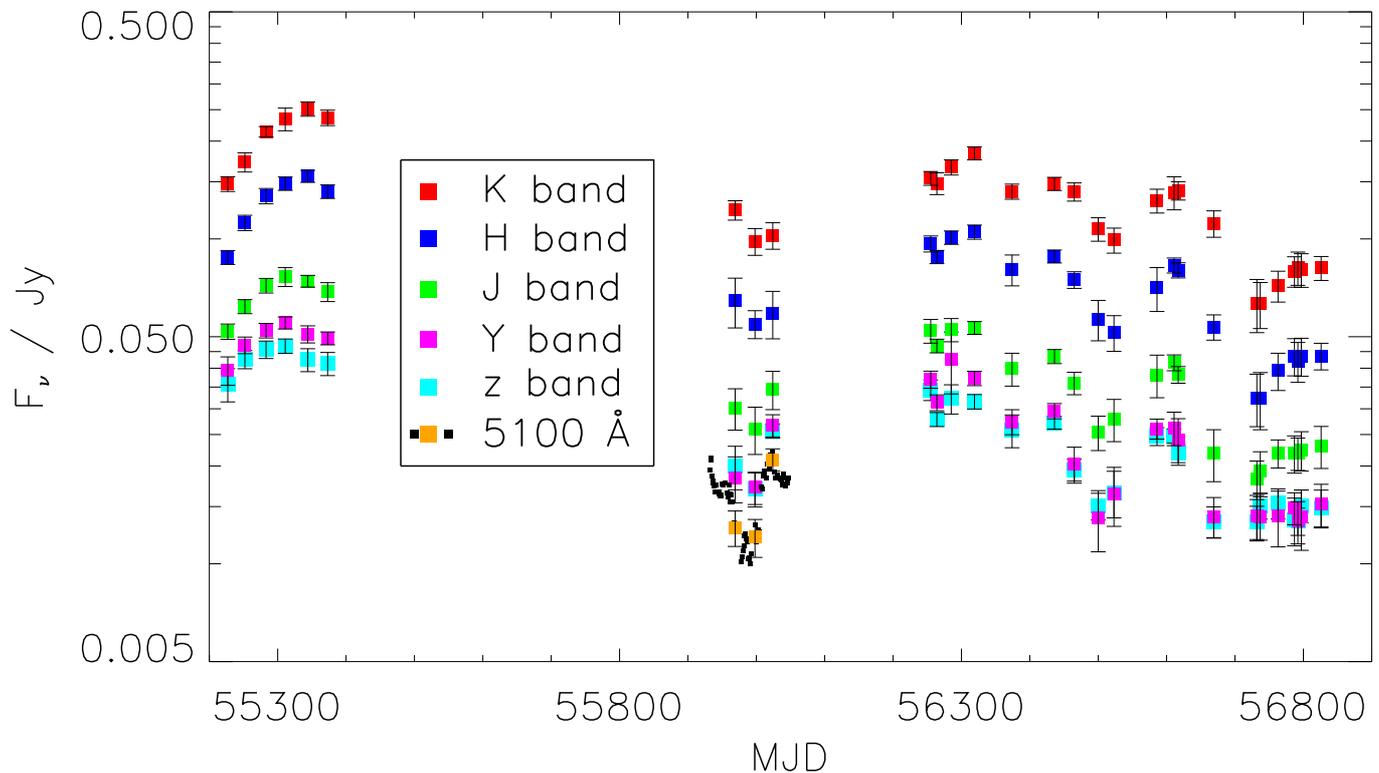}
\caption{\label{fig:fluxes} 
Calibrated 2010 - 2013
photometry of NGC4151. 
The optical data are plotted in orange for the epochs that coincide with our NIR epochs, and in black otherwise. Photometry of the NIR data was done with {\tt ISIS}, and is highly consistent with our previously determined fluxes with {\tt GALFIT} (also see Fig.~\ref{fig:ISISvsGalfit}).
}
\end{figure*}

Furthermore, we used mid-infrared (MIR) images of NGC 4151 in the WISE W1, W2, and W3 bands, downloaded from the NASA / IPAC Infrared Science Archive (IRSA), and extracted the nuclear fluxes with {\tt GALFIT}. The downloaded images in each band are a stack of 31 individual frames, taken on 2010 May 31 and 2010 December 8. Although images are only available in stacked form, the total magnitude per band is given in the ALLWISE multiepoch catalog table for each of the 31 individual observations. From a weighted magnitude difference between 2010 May and 2010 December (that we attribute to the variability of the nucleus), we derived a correction term to scale our nuclear fluxes determined with {\tt GALFIT} to the epoch 2010 May 31 (MJD 55347, then almost coinciding with Epoch 5 of our NIR observations). Our derived AGN and host fluxes are listed in Table \ref{tab:WISE}. The total flux in each band agrees very well with the average total flux for NGC 4151 given in the ALLWISE catalog.\\

\subsection{Optical data}

\begin{table*}
\caption{Optical continuum fluxes of the nucleus of NGC 4151 at 5100 \AA\ .}       
\label{tab:5100}      
\centering          
\begin{tabular}{c | c | c | c | c | c }     
\hline       
MJD & $F_\lambda$ & $\Delta F_\lambda$ & MJD & $F_\lambda$ & $\Delta F_\lambda$\\ 
 & ($10^{-13} W\mbox{m}^{-2}\mu\mbox{m}^{-1}$) & ($10^{-13} W\mbox{m}^{-2}\mu\mbox{m}^{-1}$) &  & ($10^{-13} W\mbox{m}^{-2}\mu\mbox{m}^{-1}$) & ($10^{-13} W\mbox{m}^{-2}\mu\mbox{m}^{-1}$)\\ 
\hline                 
 55932  &    2.243    &  0.208 & 55991    &  1.155    &  0.204\\
55933   &   2.401    &  0.209 & 55992    &  1.241    &  0.204 \\
 55934   &   2.443    &  0.209 & 55997    &  1.436    &  0.205 \\
55935   &   2.150    &  0.208 & 55998    &  1.392    &  0.205 \\
 55936  &   2.079    &  0.208 &  55999    &  1.515    &  0.205 \\
55937   &   2.047     & 0.207 & 56000    &  1.352    &  0.205 \\
 55938  &   1.997    & 0.207 & 56001    &  1.386    &  0.205 \\
55940   &   1.917   &   0.207 & 56002    &  1.412    &  0.205 \\
 55941   &   2.022    &  0.207 & 56003    &  1.469    &  0.205 \\
55945   &   1.914   &   0.207 & 56004    &  1.446    &  0.205 \\
 55946    &  1.885    &  0.207 & 56008    &  1.993    &  0.207 \\
55947   &   1.860    &  0.207 & 56009    &  1.955   &   0.207 \\
 55948   &   1.881    &  0.207 & 56010    &  2.147   &   0.208 \\ 
55949   &   1.860    &  0.207 & 56011    &  2.218   &   0.208 \\
 55950   &   2.035   &   0.207 & 56012    &  2.151   &   0.208 \\
55952    &  2.022   &   0.207 & 56014    &  2.116   &   0.208 \\
 55954   &   2.049   &   0.207 & 56016    &  2.342   &   0.209 \\
55957   &   1.911   &   0.207 & 56017    &  2.417   &   0.209 \\
 55958   &   1.847   &   0.207 & 56018    &  2.411   &   0.209 \\
55959    &  1.863    &  0.207 & 56019    &  2.258   &   0.209 \\
 55960     & 2.012   &   0.207 & 56021    &  2.483   &   0.209 \\ 
55961    & 1.886   &   0.207 & 56022    &  2.489   &   0.209 \\
 55962  &    1.792   &   0.206 & 56023    &  2.548   &   0.210 \\ 
55963    &  1.879   &   0.207 & 56024    &  2.413   &   0.190 \\
 55964    &  1.785   &   0.206 & 56026    &  2.205   &   0.189\\
55968    &  1.543   &   0.205 &  56028    &  2.118   &   0.189 \\
 55969    &  1.490   &   0.205 & 56031    &  2.161   &   0.189 \\
55978    &  1.171    &  0.204 & 56034    &  2.118   &   0.189 \\
 55979    &  1.208    &  0.204 & 56035    &  2.104   &   0.189 \\
55981    &  1.265   &  0.204 & 56036    &  2.087   &   0.189 \\
 55982    &  1.310   &  0.205 & 56038    &  2.018   &   0.188 \\
55983    &  1.409    &  0.205 & 56039    &  2.180   &   0.189 \\
 55984    &  1.422    &  0.205 & 56040    &  2.032   &   0.188 \\
55985    &  1.373    &  0.205 & 56042    &  2.124   &   0.189\\ 
 55987    &  1.344    &  0.205 & 56043    &  2.007   &   0.188\\ 
55988    &  1.191    &  0.204 &  56045    &  2.051   &   0.188 \\
 55989    &  1.197    &  0.204 & 56046    &  2.070   &   0.189 \\
55990    &  1.203    &  0.204 & 56047    &  2.120   &   0.189\\  
\hline              
\end{tabular}
\end{table*}

Furthermore, optical spectra were obtained between 2012 January and April
as part of a reverberation mapping campaign described
elsewhere \citep{DeRosa2014}. Observations were made on the
1.3-m McGraw-Hill Telescope at the MDM Observatory on Kitt Peak
with the Boller \& Chivens CCDS spectrograph. A 350 lines mm$^{-1}$
grating yielded a dispersion of 1.33 \AA\ pixel$^{-1}$. The
entrance slit oriented north--south (${\rm P.A.} = 0^o$) with
a projected width of $5\farcs0$. This configuration yielded
spectra covering the range 4400--5850\,\AA\ with a spectral resolution of 
7.9\,\AA. We used an extraction window of $15\farcs0$ in the
cross-dispersion direction.\\
Spectra were scaled to a common [O {\sc iii}]$\,\lambda4959$ flux 
of $3.76 \times 10^{-12}\,{\rm ergs\ s}^{-1}\,{\rm cm}^{-2}$.
The starlight contribution to each spectrum was estimated from
host-galaxy surface brightness models based on
{\em Hubble Space Telescope} images as 
$18.4\ (\pm 1.8)\ \times 10^{-15}\ {\rm ergs\ s}^{-1}\ {\rm cm}^{-2}\ {\rm \AA}^{-1}$
\citep{Bentz2013}.\\
The continuum fluxes at 5100 \AA\ (listed in Table \ref{tab:5100}) are then derived by averaging the spectrum over a 20 \AA\ range
around the lowest point, which is located between the [O {\sc iii}]$\,\lambda4959, \lambda5007$ lines and
the Fe {\sc ii} blends.\\


\section{Methods}
\label{sec:Methods}
\subsection{Dust reverberation}
\label{subsec: Dust reverberation}
The hot dust around the central nuclear region absorbs radiation emitted by the AD
and re-emits it in the infrared. This dust emission can be roughly approximated by a blackbody function. Owing to the absorption and re-emission of incident radiation by the dust, variations in the AD flux can also be detected in the dust emission with a characteristic
lag time of $\tau=R_{dust}/c$, where $R_{dust}$ is the radial distance of the hot dust from the AD and $c$ the speed of
light. 
With dust reverberation, that is to say, by monitoring the hot dust signal as it responds to the AD
signal, we determine the reverberation lag 
time and the temperature evolution of the hot dust around NGC 4151, to
infer the stability of the hot dust and to conclude whether it is close to sublimation.\\
We note that the dust
sublimation temperature depends strongly on the assumed dust species,
composition and grain sizes, all of which might change significantly with
radial
distance from the source. Without detailed knowledge of these  dust
properties,
it is therefore not possible to determine whether the innermost dust
is close to sublimation from measuring the dust
temperature in one single epoch. Monitoring the hot dust temperature for an enhanced period, however, allows us to draw more reliable
conclusions about the immediate dust state. If we observe rising temperatures following increased
AD
emission, we can conclude that the hot dust has not yet reached its sublimation temperature. Clearly, the temperature evolution is a more robust indicator than the absolute dust temperature in one single epoch.\\

To be able to measure the dust response to varying AD emission, we decompose the total 
fluxes in each epoch and band as the sum of an AD and a blackbody component that we attribute to the dust. In contrast to the single-epoch SED fits (Eq.~\ref{eq:Model_single}) used in Paper I, we perform a multiepoch, multiwavelength fit on our complete data set, of the form 
\begin{equation}
F_{\lambda}(\lambda,t,\mathbf{x})=C_1\cdot F_z(t)\cdot \lambda^{-\alpha}+BB(C_2,\lambda,T(v\cdot F_z(t-\tau)))\, ,
\label{eq:Model}
\end{equation} 
Here, $\mathbf{x}$ denotes our complete vector of model parameters. The first term on the righthand side of Eq.~\ref{eq:Model} refers to the AD emission, usually described as a power law in terms of $\lambda$, with a power-law index $\alpha$ which we assume does not change significantly over the observed flux range. For completeness, however, we also test for a varying $\alpha$ and its influence on our results, see Sects.~\ref{subsec:alpha_var} and \ref{subsec:VarLag}. $C_1$ is a proportionality constant. In the second term on the righthand side of Eq.~\ref{eq:Model}, $BB(.)$ denotes the blackbody function
\begin{equation}
BB(C_2,\lambda,T)=C_2\cdot\frac{2hc^2}{\lambda
^5}\frac{1}{\mbox{e}^{hc/\lambda
k_B T}-1} \, .
\label{eq:blackbody}
\end{equation} 
with $T$ the blackbody temperature of the hot dust and $C_2$ the blackbody constant (consisting of a mixture of emissivity, solid angle,
and surface filling factor). In Eq.~\ref{eq:Model}, we assume that the AD signal has the shape of the interpolated $z$-band signal, which is justified because the $z$ band is known to be dominated by AD emission \citep{Rif2009}. Starting from an initial temperature $T_0$, the blackbody temperature $T$ evolves by construction, for $t > \tau$, in response to the AD variations as 
\begin{equation}
\mbox{d}T(t)/T(t)=v\cdot\frac{1}{4}\cdot\mbox{d}L_z(t-\tau)/L_z(t-\tau)\, .
\label{eq:T}
\end{equation} 
The variability factor $v$ accounts for the possibility that the $z$-band variability is not completely reprocessed by the hot dust \citep{Hoe2011}, which would lead to values of $v < 1$. Furthermore, values of $v \neq 1$, might indicate that the variability of the heating signal is underestimated or overestimated by the $z$-band variability. The hot dust around AGN is heated mainly by the UV part of the AD radiation. We take the $z$-band flux as a proxy for this UV heating; however, the actual UV radiation is probably even more variable.
Our full parameter vector thus comprises $\mathbf{x}=(C_1,\alpha,C_2,T_0,\tau,v)$. Our new multiepoch approach Eq.~\ref{eq:Model}-\ref{eq:T}, in particular the temperature evolution given by Eq.~\ref{eq:T}, is justified by the measured hot dust temperature evolution
obtained from preceding single-epoch SED fits of the form
\begin{equation}
F_{\lambda}(\lambda,\mathbf{x})=\widetilde{C}_1\cdot\lambda^{-\alpha}+C_2\cdot\frac{2hc^2}{\lambda
^5}\frac{1}{\mbox{e}^{hc/\lambda
k_B T}-1} \, ,
\label{eq:Model_single}
\end{equation}
i.e.~AD plus blackbody contribution. These single-epoch fits were used in Paper I, and were also performed for each epoch of the complete 2010-2014 data set, prior to the multiepoch approach described by Eqs.~\ref{eq:Model}-\ref{eq:T}. 
As can be seen in Fig.~\ref{fig:LzAD_vs_T1}, the blackbody temperature obtained from the single-epoch SED fits tracks the 
AD signal closely, with a stable delay of roughly 30 days, and the temperature variation is significant, thus justifying our new model.  Furthermore, behavior as in Eq.~\ref{eq:T} is expected according to \citet{Hoe2011}. Compared to Eq.~\ref{eq:Model_single}, the data are temporally connected in our new approach. Using the temperature evolution described by Eq.~\ref{eq:T} drastically reduces the number of parameters of the multiepoch fit, since we only need to estimate the initial blackbody temperature $T_0$ instead of a temperature for each single epoch. This makes the algorithm more robust against degeneracies between $T$ and $C_2$.\\
To check whether the distribution of the innermost hot dust is fairly compact or, in contrast, substantially radially extended (as we conjectured in Paper I), we tested for the possibility of a further blackbody component contributing to the hot dust signal. Therefore, in addition to the 1BB-model of Eq.~\ref{eq:Model}, we fitted a 2BB-model of similar type to our data:
\begin{eqnarray}
F_{\lambda}(\lambda,t,\mathbf{x})&=&C_1\cdot F_z(t)\cdot \lambda^{-\alpha}+BB1(C_2,\lambda,T_1(v_1\cdot F_z(t-\tau_1))) \nonumber \\
& &+BB2(C_3,\lambda,T_2(v_2\cdot F_z(t-\tau_2)))\, 
\label{eq:Model2}
\end{eqnarray} 
with the temperature evolution of the second blackbody given by an expression as in Eq.~\ref{eq:T}, though with different starting temperature, time lag, and variability factor. This 2BB-model has four additional parameters, and $\mathbf{x}=(C_1,\alpha,C_2,T_{0,1},\tau_1,v_1,C_3,T_{0,2},\tau_2,v_2)$.

\subsection{Interpolation of the AD signal}
\label{subsec:interp}
\begin{figure}
\includegraphics[width=9.cm,angle=0,clip=true]
{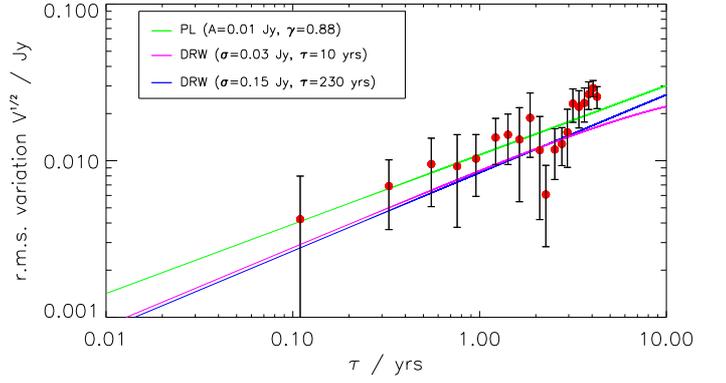}
\caption{\label{fig:strucfun}
Estimated $(v_{ij})^{1/2}$ (red dots) from the $z$ band data versus $\tau_{ij}$, shown on logarithmic scale. The three different structure function models that we fit to these data (see text of Sect.~\ref{subsec:interp}) are represented by the green, magenta, and blue lines.}
\end{figure}

To calculate the blackbody temperature $T$ described by Eq.~\ref{eq:T} of our new approach for any arbitrary value of $\tau$ in the range of the data, we first need to interpolate our input accretion disk signal, i.e.~the $z$-band signal. This is done with the method for interpolation, realization, and reconstruction of noisy, irregularly sampled data by \citet{Ryb1992} and \citet{Pre1992}, which was previously used for reverberation studies of the BLR, for example by \citet{Zu2011} or \citet{Hern2014}. The AD signal can be described as a stochastic 
process characterized by two structure function parameters, which are estimated from the data. With this we can predict 
the signal for unmeasured times.\\
Our $M=29$ measurements $\mathbf{y}=\{y_i\}, i=1,...,M$ of the $z$-band flux can be written as
\begin{equation}
\mathbf{y}=\mathbf{s}+\mathbf{E}\bar{y}+\mathbf{n}\, ,
\end{equation}
where $\mathbf{s}$ represents the intrinsic variability signal, $\mathbf{E}$ is the vector $(1,1,1...,1)^T$, $\bar{y}$ is an appropriate estimator of the mean of the data, and  $\mathbf{n}$ represents the measurement noise. The intrinsic signal $\mathbf{s}$ can be described as a stochastic process with a certain correlation structure,
and we can calculate an estimate $\hat{s_*}$ of the signal at any (measured or unmeasured) point using the already obtained $M$ measurements:
\begin{equation}
\hat{s_*}=\mathbf{\hat{d}}_*^T (\mathbf{y}-\mathbf{E}\bar{y}) + x_*\ .
\label{eq:est}
\end{equation}
Here, $\mathbf{\hat{d}}_*$ are linear coefficients that depend on the particular point to be estimated, and $x_*$ is the discrepancy between the estimated and the true value. Minimizing the discrepancy with respect to the linear coefficients then yields the solution for the coefficients, and the least variance estimate of our signal 
\begin{equation}
\hat{s_*}=\mathbf{S}_*^T [\mathbf{S}+\mathbf{N}]^{-1} (\mathbf{y}-\mathbf{E}\bar{y}) +\bar{y}\, .
\label{eq:leastvar}
\end{equation}
Here, $\mathbf{S}$ is the covariance matrix of the data, $\mathbf{S}_*$ is the covariance vector between the measured data points and the new data point, and $\mathbf{N}$ is the noise covariance matrix.
In Eqs.~\ref{eq:est} and \ref{eq:leastvar}, the mean $\bar{y}$ is subtracted from the data before determining the coefficients \footnote{The appropriate mean is calculated by 
\begin{equation*}
\bar{y}=\frac{\mathbf{E}^T[\mathbf{S}+\mathbf{N}]^{-1}\mathbf{y}}{\mathbf{E}^T[\mathbf{S}+\mathbf{N}]^{-1}\mathbf{E}}\, ,
\end{equation*} and corresponds to the value $\bar{y}$ that minimizes 
\begin{equation*}
\chi^2=(\mathbf{y}-\mathbf{E}\bar{y})^T [\mathbf{S}+\mathbf{N}]^{-1} (\mathbf{y}-\mathbf{E}\bar{y})
\end{equation*} when subtracted from the data.} and is added back to the estimate afterward.
The variance of the true value $s_*$ about the best estimate $\hat{s_*}$ is given by
\begin{equation}
\langle(s_*-\hat{s_*})^2\rangle=\langle s_*^2\rangle - \mathbf{S}_*^T [\mathbf{S}+\mathbf{N}]^{-1}\mathbf{S}_*\, .
\end{equation}

\begin{figure}
\includegraphics[width=9.cm,angle=0,clip=true]
{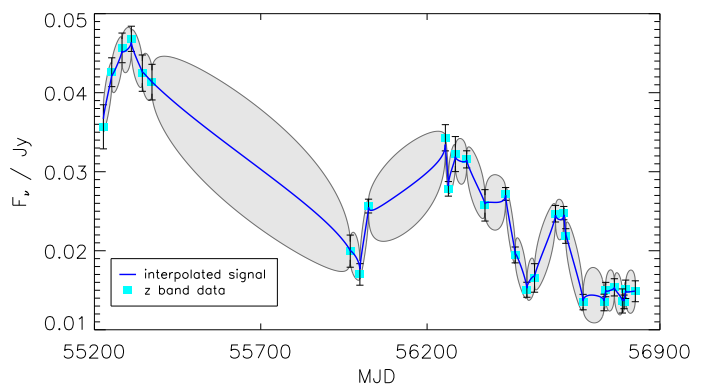}
\caption{\label{fig:Lz_Fnu_ISIS_interpolated}
$z$-band light curve, interpolated with the method of \citet{Ryb1992} and \citet{Pre1992}. We interpolated using the three models described in Sect.~\ref{subsec:interp}. For clarity, only the interpolation resulting from the power-law structure function is shown here. This interpolated signal is the least-variance estimate of the stochastic process and therefore very smooth. The error of this prediction is indicated by the shaded gray region. Single realizations of the process have much more structure on short timescales and will also make excursions outside of the estimated error regions.}
\end{figure}

To apply Eq.~\ref{eq:leastvar}, we need to estimate the covariance structure of the underlying stochastic process from our data. Assuming stationarity, so that $S_{ij}\equiv S(t_i-t_j)\equiv S(\tau)$, the covariance matrix can be replaced by the population mean square and the structure function $V(\tau)$: 
\begin{equation}
S(\tau)=\langle s^2\rangle-V(\tau)\, ,
\end{equation}
\begin{equation}
V(\tau)\equiv \frac{1}{2}\langle[s(t+\tau)-s(t)]^2\rangle\, .
\end{equation}
Following the method described by \citet{Pre1992}, we calculate a lag $\tau_{ij}=|t_i-t_j|$ and an estimate of the structure function $v_{ij}=(y_i-y_j)^2-n_i^2-n_j^2$ for each pair $(i,j)$, where $n_i$ is the measurement error of data point $y_i$. The $M(M-1)/2 = 406$ pairs are binned into 20 bins, equally spaced in $\tau$. In Fig.~\ref{fig:strucfun}, $\sqrt{v_{ij}}$ is plotted against $\tau_{ij}$ on a logarithmic scale. We fit a straight line 
\begin{equation}
\mbox{log}(\sqrt{V(\tau)})=\mbox{log}(A)+\frac{\gamma}{2}\mbox{log}(\tau)
\end{equation} 
to these data, representing a power-law model of the form 
\begin{equation}
V(\tau)=A^2 \cdot \left(\frac{\tau}{\mbox{1 yr}}\right)^{\gamma}\, ,
\end{equation} 
where $A$ is the average flux variability on a one-year timescale, and $\gamma$ is the gradient of this variability.
For the structure function parameters, we obtain best fit values of $A=0.011\pm0.001$ Jy (resp.~$A=0.423\pm0.019 \mbox{W m}^{-2}\mu\mbox{m}^{-1}$) and $\gamma=0.881\pm0.067$. With these values, we calculate the covariance matrix and vector, and interpolate our $z$-band light curve according to Eq.~\ref{eq:leastvar}. The interpolated $z$-band light curve is shown in Fig.~\ref{fig:Lz_Fnu_ISIS_interpolated}.
Since our measurement errors are generally small (at the 5\% level), the interpolated light curve runs almost perfectly through our data points. In Fig.~\ref{fig:strucfun}, a potential substructure seems to be evident around $\tau=2.5$ yr. However, this apparent feature might be merely an artifact, caused by insufficient sampling in that range of $\tau$. Indeed, the bins around $\tau=2.5$ yrs contain by far the least number of data points, obviously owing to the large data gap between June 2010 and February 2012. An apparent feature seen in Fig.~\ref{fig:strucfun} seems to be common in the empirical structure functions of single targets (see e.g.~figures in \citet{Pre1992,Schm2010,Morg2014}), while it is averaged out in the observed structure function of samples of AGNs (see e.g.~\cite{Schm2010}).\\
Our derived structure function parameters agree with values found in the literature. While for quasars, typical values of the power-law slope are $\gamma\approx0.3-0.4$ \citep{Bau2009,Schm2010}, the structure function of Seyferts is generally found to be steeper \citep{Hawk2002}, except for very short timescales, which are dominated by noise. Indeed, specifically for NGC 4151, \citet{Czer2003} report slopes of $\approx0.65-1.0$ for the $V$ band structure function on different timescales and in different epochs.\\

An alternative approach of the structure function to describe AGN variability is the damped random walk (DRW) model:
\begin{equation}
V(\tau)=\sigma^2\left(1-\mbox{exp}(-\tau/\tau_d)\right),
\end{equation}
where $\sigma^2$ is the long-term variance of the process, and $\tau_d$ the damping timescale.
While for a substantial $\tau$ range the variability increases with increasing time lag following a power law $V(\tau)\propto\tau^{\gamma}$, the DRW model considers that there is a plateau where the variability saturates, at time lags that are longer than the longest correlation timescale of the stochastic process. When we fit this model to our observed structure function for the NGC 4151 $z$ band signal, we obtain values $\sigma=0.148\pm0.505$ Jy and $\tau_d=230.0\pm576.1$ yrs. Clearly, the fitted damping timescale, typically between 0.1 and 3 years \citep{MacL2010}, seems physically unreasonable. When looking at the observed structure function data plotted in Fig.~\ref{fig:strucfun}, we can see that our time series is obviously not long enough to sufficiently constrain the damping timescale for NGC 4151, as also evident from its large error. Obviously, no plateau is reached, but the variability increases throughout the observed time range. This is consistent with the results of \citet{Czer2003}, who find an unusually long damping timescale of roughly ten years for this object. We performed the fit again with fixed $\tau_d=10$ yrs, leading to $\sigma=0.032\pm0.002$ Jy.\\
The obtained structure functions for the power law and the DRW model are plotted in Fig.~\ref{fig:strucfun}. To test the influence of the particular choice of structure function on the robustness of our results, we performed all our following analyses (i.e., the interpolation of the AD signal as well as the various fits presented in Sects.~\ref{subsec:alpha_var} and \ref{sec:Discussion}) with all three models shown in Fig.~\ref{fig:strucfun}.

\subsection{Model fitting using a DE-MC algorithm}

\underline{Differential evolution Markov chain}\\
Compared to Paper I, in which we used the routine {\tt mpfit} to fit the model of interest to our data, we now infer the best model parameters using a differential evolution Markov chain  (DE-MC) algorithm, which was proposed by \citet{Braak2006}. DE-MC is a population MCMC algorithm, with multiple chains running in parallel. 
Compared to random walk metropolis (RWM) type MCMC algorithms, the main advantage of DE-MC is that the problem of choosing a jumping distribution, hence an appropriate scale and direction of the jumps, is solved by generating jumps as a fixed multiple of two randomly selected parameter vectors $\mathbf{x}_{R1}$ and $\mathbf{x}_{R2}$ that are currently in the population:
\begin{equation}
\mathbf{x}_p=\mathbf{x}_i+\gamma\cdot(\mathbf{x}_{R1}-\mathbf{x}_{R2})+\mathbf{e}
\label{eq:DEMC}
\end{equation}
where $\gamma > 0$, and the vector  $\mathbf{e}$ of small random numbers is added to ensure that the whole parameter space can be reached. These random numbers are drawn from a symmetric (e.g.~Gaussian) distribution that has small
variance compared to the target variance. 
In the proposal scheme of Eq.~\ref{eq:DEMC}, 
the chains learn from one another, hence the name ``differential evolution''.\\


\begin{figure}[htb]
   \centering
  \includegraphics[width=8.cm]{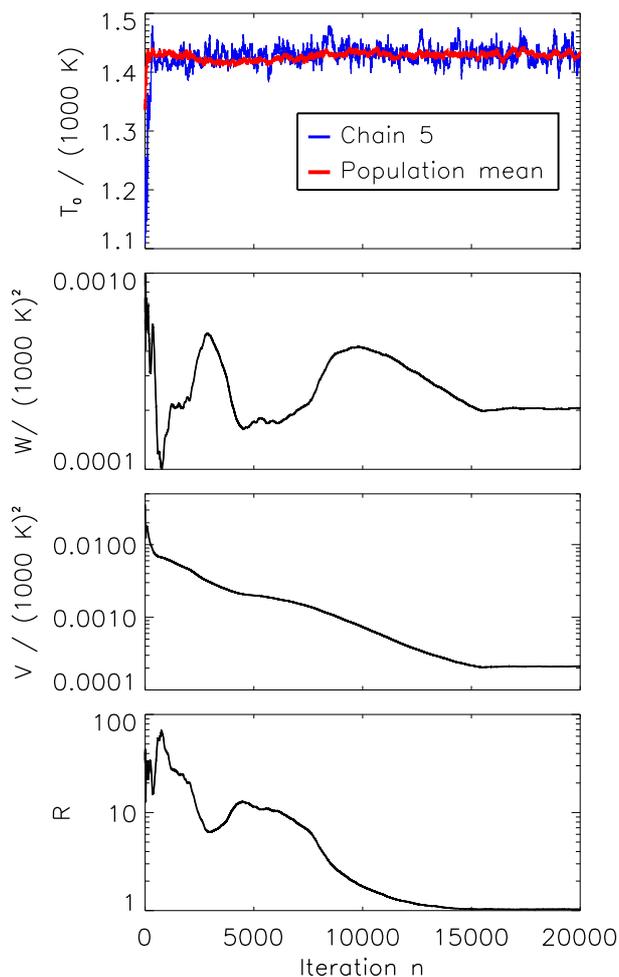}
   \caption{\label{fig:Gelman} Evolution of the parameter $T_{0,1}$ (for one chain and the population mean) of the 1BB model, for one exemplary run. Also shown are the evolution of the pooled within chain variance $W$ (Eq.~\ref{eq:W}), maximum variance estimate $\hat{V}$ (see Eq.~\ref{eq:Rhat}) and the resulting convergence parameter, i.e., the PSRF $\hat{R}$ (Eq.~\ref{eq:Rhat}). Convergence was diagnosed at iteration step $n=13599$, where $\hat{R}<1.1$ was reached for all parameters.}
    \end{figure}

\begin{figure}[htb]
   \centering
  \includegraphics[width=8.cm]{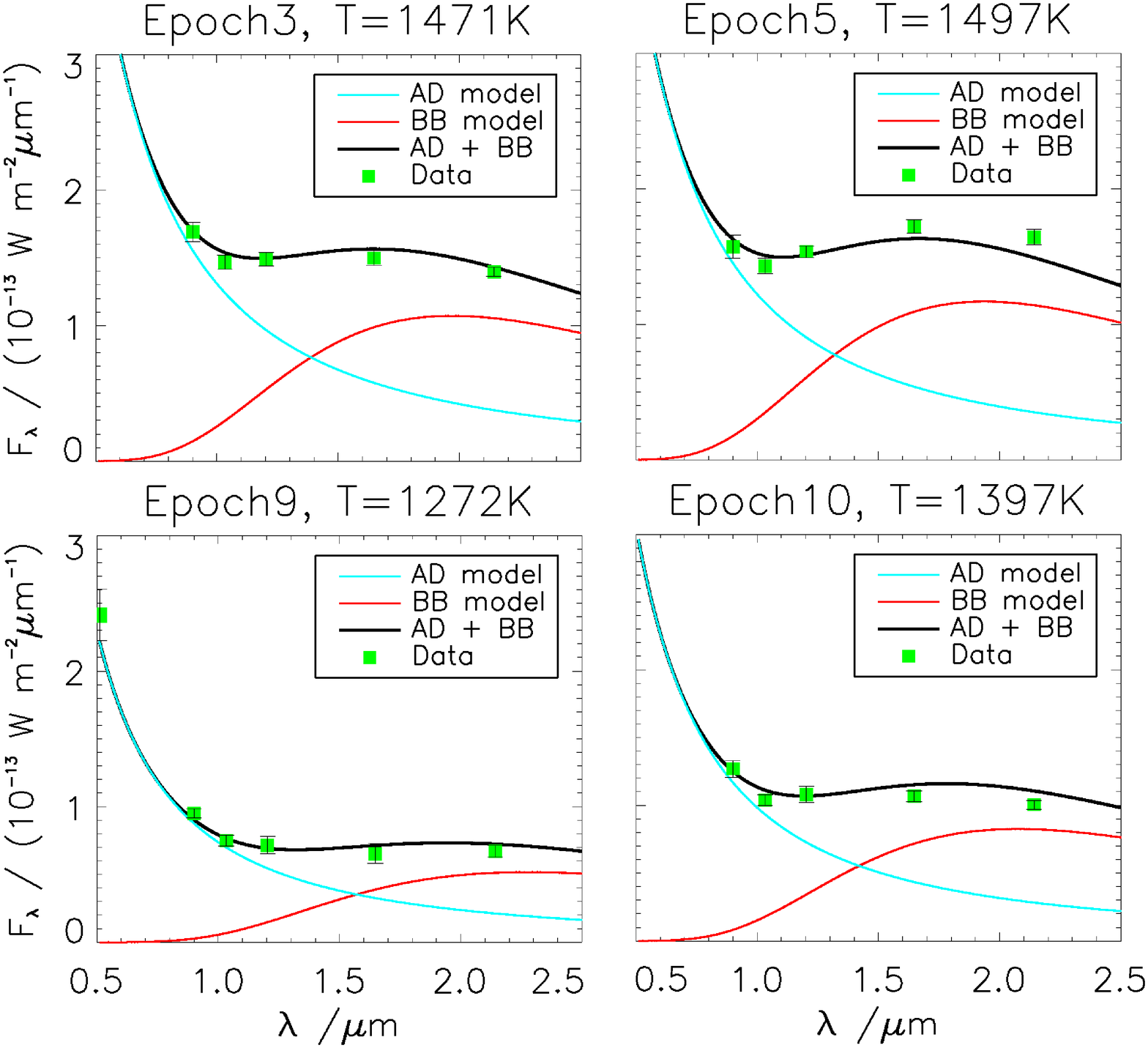}
   \caption{\label{fig:4epochs} Our data and the 1BB model fit for the epochs 3,5,9, and 10. Overplotted is the AD model, the 1BB model, and the sum of both components. From the data, we observe a clear rise in the blackbody flux, and a clear shift of the blackbody emission peak to shorter wavelengths until epoch 5. This emerging bump is confirmed by the blackbody temperature maximum in epoch 5 of our single-epoch fit (see Fig.~\ref{fig:LzAD_vs_T1}).}
    \end{figure}

\begin{table*}
\caption{Upper and lower limits $a,b$ for uniform pdfs of the model parameters, for the 1BB model (with constant and varying $\alpha$) and the 2BB model.
}       
\label{tab:priors}      
\centering          
\begin{tabular}{l | c | c | c | c | c | c | c }     
\hline       
\hspace{-0.15cm}Parameter & Unit & \multicolumn{2}{c}{1BB model} & \multicolumn{2}{c}{1BB model (var.~$\alpha$)} & \multicolumn{2}{c}{2BB model}\\
\hspace{-0.15cm} & & $a$ & $b$ & $a$ & $b$ & $a$ & $b$\\
   
\hline                 
   \hspace{-0.15cm}$C_1$ & -- & 0.0 & 0.85 & 0.0 & 0.85 & 0.0 & 0.85\\ 
\hspace{-0.15cm}$\alpha$ & -- & 1.5 & 2.5 & 1.5 & 2.5 & 1.5 & 2.5\\ 
\hspace{-0.15cm}$\alpha_1$ & -- & -- & -- & 0.0 & 1.0 & -- & --\\ 
\hspace{-0.15cm}$C_2$ & $10^{-18} \mbox{ster}$ & 1.0 & 30.0 & 1.0 & 30.0 & 1.0 & 20.0\\ 
\hspace{-0.15cm}$T_{0,1}$ & $1000 \mbox{K}$ & 1.0 & 2.0 & 1.0 & 2.0 & 1.0 & 2.0\\ 
\hspace{-0.15cm}$\tau_1$ & $\mbox{days}$ & 0.0 & 80.0 & 0.0 & 80.0 & 0.0 & 40.0\\ 
\hspace{-0.15cm}$v_1$ & -- & 0.5 & 2.0 & 0.5 & 2.0 & 0.5 & 2.0\\
\hspace{-0.15cm}$C_3$ & $10^{-18} \mbox{ster}$ & -- & -- & -- & -- & 10.0 & 100.0\\ 
\hspace{-0.15cm}$T_{0,2}$ & $1000 \mbox{K} $ & -- & -- & -- & -- & 0.5 & 1.0\\ 
\hspace{-0.15cm}$\tau_2$ & $\mbox{days}$ & -- & -- & -- & -- & 30.0 & 100.0\\ 
\hspace{-0.15cm}$v_2$ & -- & -- & -- & -- & -- & 0.5 & 2.0\\ 
\hline        
    
\end{tabular}
\end{table*}

\begin{table*}
\caption{Global mean and errors for the parameters of our various models, and reduced $\chi^2$ of the fit. The errors are given by $\Delta x=\sqrt{\hat{V}}$ for each parameter. All results shown here are derived from using the power-law structure function.
}       
\label{tab:pars}      
\centering          
\begin{tabular}{l | c | c | c | c | c | c }     
\hline       
\hspace{-0.15cm}Parameter & Unit & 1BB & 1BB (var. $\alpha$) & 1BB (ep.~1-9) & 1BB (ep.~7-29)  & 2BB \\
\hline              
   \hspace{-0.15cm}$C_1$ & -- & 0.78 $\pm$ 0.01 & 0.78 $\pm$ 0.01 & 0.78 $\pm$ 0.01 & 0.78 $\pm$ 0.01 & 0.77 $\pm$ 0.01 \\ 
\hspace{-0.15cm}$\alpha$ & -- & 1.63 $\pm$ 0.04& 1.56 $\pm$ 0.07 & 1.65 $\pm$ 0.07& 1.63 $\pm$ 0.04 & 1.65 $\pm$ 0.04\\ 
\hspace{-0.15cm}$\alpha_1$ & -- & -- & 0.10 $\pm$ 0.07& -- & -- & -- \\ 
\hspace{-0.15cm}$C_2$ & $10^{-18} \mbox{ster}$ & 3.82 $\pm$ 0.19& 3.88 $\pm$ 0.20& 4.63 $\pm$ 0.33& 3.52 $\pm$ 0.22 & 3.16 $\pm$ 0.34\\ 
\hspace{-0.15cm}$T_{0,1}$ & $1000 \mbox{K}$ & 1.436 $\pm$ 0.015& 1.436 $\pm$ 0.015 & 1.392 $\pm$ 0.020  & 1.455 $\pm$ 0.018&  1.479 $\pm$ 0.027\\ 
\hspace{-0.15cm}$\tau_1$ & $\mbox{days}$ & 31.0 $\pm$  1.6 & 30.8 $\pm$ 1.5 & 42.5 $\pm$ 4.0 & 29.6  $\pm$ 1.7 & 29.3 $\pm$ 1.9\\ 
\hspace{-0.15cm}$v_1$ & -- & 0.81 $\pm$ 0.02& 0.82 $\pm$ 0.02& 0.96 $\pm$ 0.04& 0.81 $\pm$ 0.03& 0.80 $\pm$ 0.02\\
\hspace{-0.15cm}$C_3$ & $10^{-18} \mbox{ster}$ & -- & -- & -- & -- & 56.47 $\pm$ 25.65\\ 
\hspace{-0.15cm}$T_{0,2}$ & $1000 \mbox{K} $ & -- & -- & -- & -- &  0.698 $\pm$ 0.064\\ 
\hspace{-0.15cm}$\tau_2$ & $\mbox{days}$ & -- & -- & -- & -- & 67.2 $\pm$ 8.3\\ 
\hspace{-0.15cm}$v_2$ & -- & -- & -- & -- & -- & 1.26 $\pm$ 0.20\\ 
\hspace{-0.15cm}$\chi^2_{red}$ & -- & 1.87 &  1.87 & 1.13 & 1.76 & 1.89\\ 
\hline        

\end{tabular}
\end{table*}
\underline{Implementation of the algorithm}\\

We ran the DE-MC algorithm for the six-parameter 1BB model of Eq.~\ref{eq:Model}, with 15 simultaneous Markov chains and 25 simultaneous chains for the case of the ten-parameter 2BB model of Eq.~\ref{eq:Model2}. We found $\gamma=2.38/\sqrt{8d}$ to be a good value for achieving the desired acceptance rate of roughly 25\%. In each iteration step and for each chain, the algorithm evaluates -- for the current parameter vector $\mathbf{x}_i$ and for the proposal vector $\mathbf{x_p}$ -- the posterior probability density function (pdf)
\begin{eqnarray}
p(\mathbf{x} | \mathbf{y}) & \propto & p(\mathbf{x})\cdot p(\mathbf{y} | \mathbf{x})\nonumber \\
& \propto & p(\mathbf{x})\cdot \mbox{exp}\left(-\frac{\chi^2}{2}\right)\nonumber \\
&=& p(\mathbf{x})\cdot \mbox{exp}\left(-\frac{1}{2}\sum_{k=1}^N\left(\frac{m_k-y_k}{\sigma_k}\right)^2 \right)\, ,
\label{eq:posterior}
\end{eqnarray} 
which is proportional to the prior pdf times the likelihood function according to Bayes' theorem. Here, $\mathbf{x}$ represents the parameter vector, and $\mathbf{y}$ are the data. The notations $m_k$ and $y_k$ respectively refer to the values of the model and the data for the $k$th of the $N=211$ NIR plus optical data points (the WISE data were not included in the model fit), and $\sigma_k$ is the related photometric error. To not overweight the low flux epochs around MJD 56000, where all of the 5100 \AA\ measurements are assembled, we decided to use a weighted $\chi^2=\chi^2_{NIR}+\chi^2_{5100}/3$. Thus, we obtain an effective number of $N=160$ data points.\\
For each parameter $x$, we chose a uniform prior pdf
\begin{equation}
p(x)=\left\{\begin{array}{ll} \frac{1}{b-a}, & a \leq x \leq b \\
\newline
\\
         0, & x < a \, \vee \, x > b\end{array} \right .
\end{equation}
within the limits $a,b$ as given in Table \ref{tab:priors}, in order to exclude unphysical parameter ranges. The posterior pdf is then directly proportional to the likelihood, and the maximum of the posterior occurs when $\chi^2$ is smallest. The upper limit for $C_1$ is given by the fact that for the $z$ band ($\lambda\approx 0.9 \mu m$), $C_1\cdot\lambda^{\alpha}$ must not exceed $1$. In the 2BB model, the maximum temperature for the second blackbody component is given by $T_{max}=T_1\cdot\sqrt{\tau_1/\tau_2}$, corresponding to the case that all radiation from the AD can reach the second blackbody, i.e. the covering factor of the first blackbody is negligible. We chose $\tau_2=100$ days as an upper limit for the reverberation lag of the second blackbody component, because at that distance (and resulting low temperatures), its contribution to the NIR fluxes would become negligible, unless the blackbody constant $C_3$ would be several orders of magnitude higher than $C_2$. We further set $p(\tau_2)=0$ for $\tau_2 < \tau_1$.\\
As initial distribution for the parameters, we took a uniform distribution within the limits of our priors, which is thus sufficiently overdispersed. \\
For monitoring convergence of the DE-MC algorithm, we used the convergence criteria proposed by \citet{Gel1992} and \citet{Gel1998}, i.e., the $\hat R$ value, and visual inspection of $W$ and $\hat{V}$, as explained in Sect.~\ref{sec:Gelman}. 
The 1BB model converged within 10000-15000 iterations on average. We considered $\hat R \leq 1.1$ as close enough to 1, \citet{Braak2006} uses $\hat R \leq 1.2$. Figure \ref{fig:Gelman} shows the evolution of the parameter $T_{0,1}$ (which took longest to converge) for one exemplary run, as well as the evolution of $\hat R$, $\hat V$ and $W$ for  this parameter. Visual inspection shows that $W$ and $\hat{V}$ are not evolving anymore and convergence was diagnosed correctly. 
The alternative ten-parameter 2BB model took 50000-100000 iterations on average to converge.

%
%
%

\section{Results}
\label{sec:Results}
\subsection{Constant power-law index}

All results presented in this section were obtained using the power-law structure function model and a constant AD power-law index $\alpha$. For results under the use of different structure function and $\alpha$ models, see Sects.~\ref{subsec:alpha_var} and \ref{subsec:VarLag}.
The temporal evolution of the photometry of the nucleus of
NGC 4151 is shown in Fig.~\ref{fig:fluxes}. The flux variations in the
other
bands show a time lag behind those of the AD-dominated $z$ band -- except for the 5100 \AA\ fluxes, which seem to be concurrent with the $z$-band fluxes --  and this lag increases with
wavelength. This band-dependent behavior can be explained by the reverberation delay of the hot dust and by the increasing dust contribution and decreasing AD contribution for longer
wavelengths.\\
\begin{figure*}[htb]
   \centering
  \includegraphics[width=17.cm]{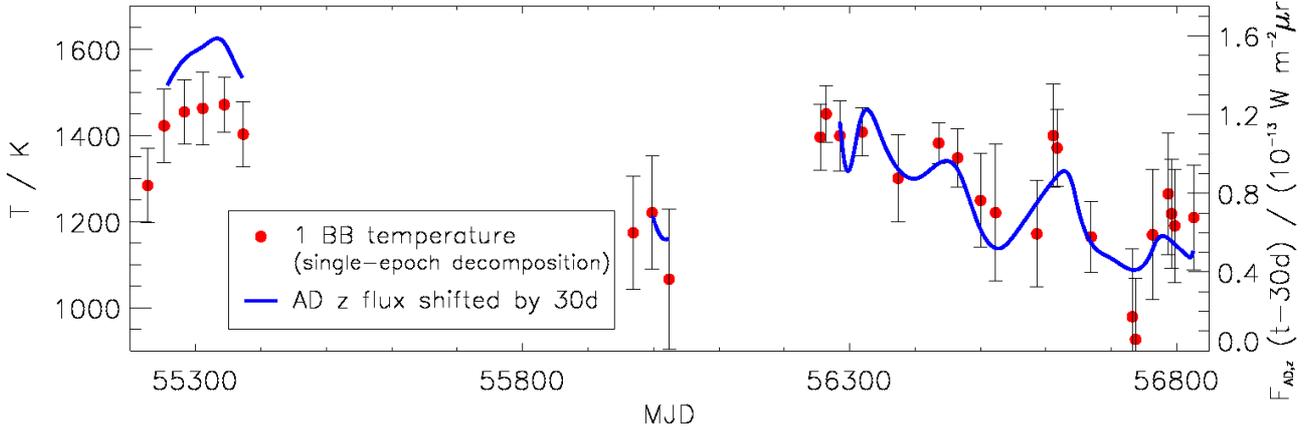}
   \caption{\label{fig:LzAD_vs_T1} Hot dust temperature evolution derived from the single-epoch decomposition. The hot-dust temperature changes follow the relative AD $z$-band flux changes with a delay of roughly 30 days.}
    \end{figure*}

\begin{figure*}[htb]
   \centering
  \includegraphics[width=18.cm]{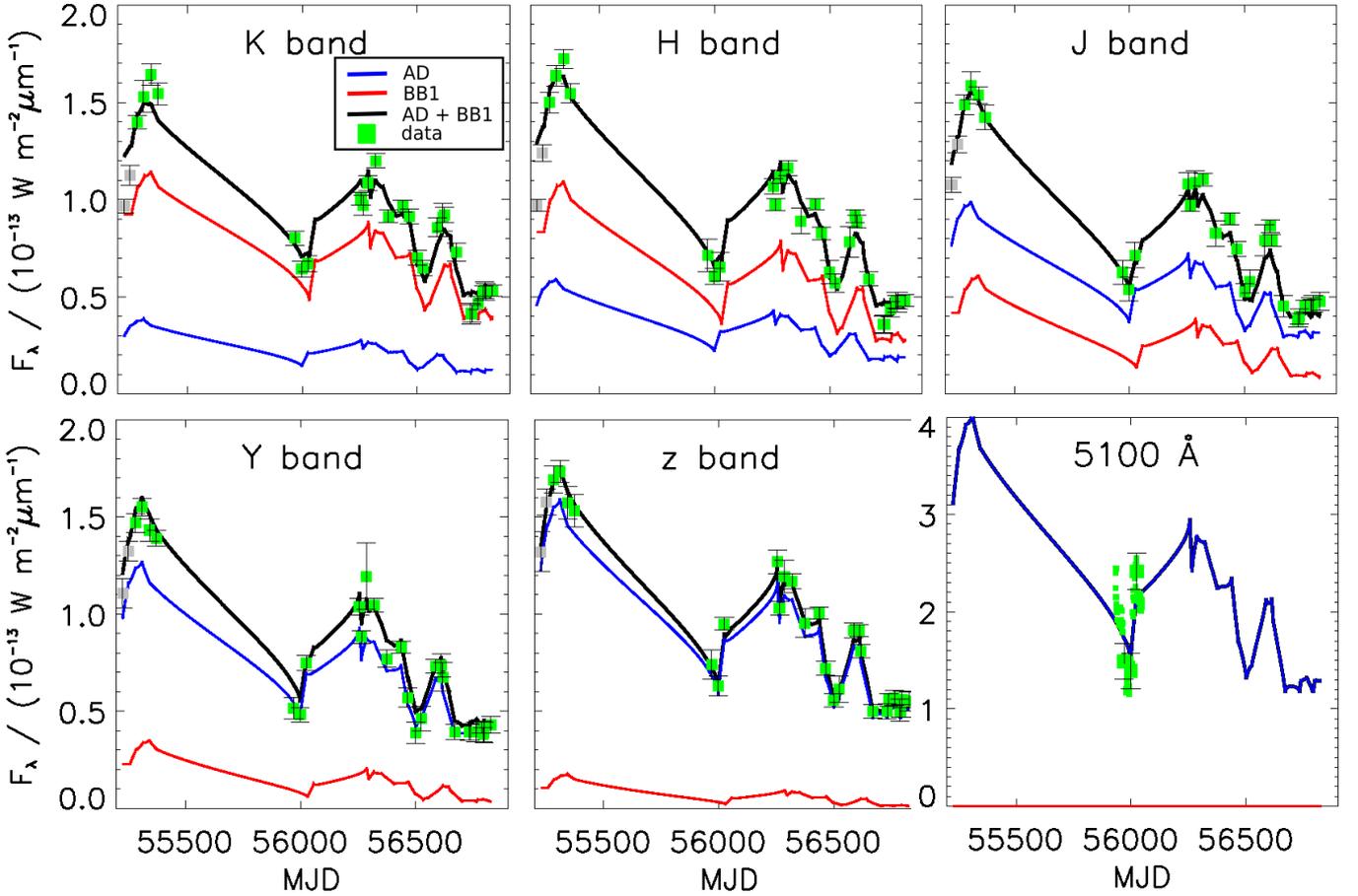}
   \caption{\label{fig:Fit_1BB} Results of our 1BB model fit. Plotted are the $K,H,J,Y,z$, and 5100 \AA\ data (green) over time. The bigger dots with errors bars in the 5100 \AA\ panel mark the data points that coincide with our NIR epochs. The red line represents the blackbody contribution for each band, blue the AD contribution, and the black line is the sum of both. Epochs 1 and 2 are not included in the evaluation of the fit (as $t < \tau$).}
    \end{figure*}

Figure \ref{fig:LzAD_vs_T1} shows a single-epoch decomposition of the fluxes into AD and hot dust according to Eq.~\ref{eq:Model_single}. The hot dust temperature closely follows the overplotted AD $z$-band flux with a delay of $\approx$ 30 days, which justifies the model described by Eqs.~\ref{eq:Model}-\ref{eq:T}. The minimum that is seen in the temperature around MJD 56700 (and also in the corresponding $JHK$ flux minima) is not covered within the sampling of the $z$-band data (see Fig.~\ref{fig:Fit_1BB} and Fig.~\ref{fig:fluxes}), therefore we observe a substantial discrepancy here.
Comparing epochs 1-9 with epochs 10-29, one can see that the ratio $F_{AD,z} / T$ is higher in the first epochs, already indicating our later finding that in these early epochs the reverberation lag might be larger than in the later epochs, so that the AD flux is more diluted and the dust heated less in epochs 1-9.\\

\begin{figure*}[htb]
   \centering
  \includegraphics[width=18.cm]{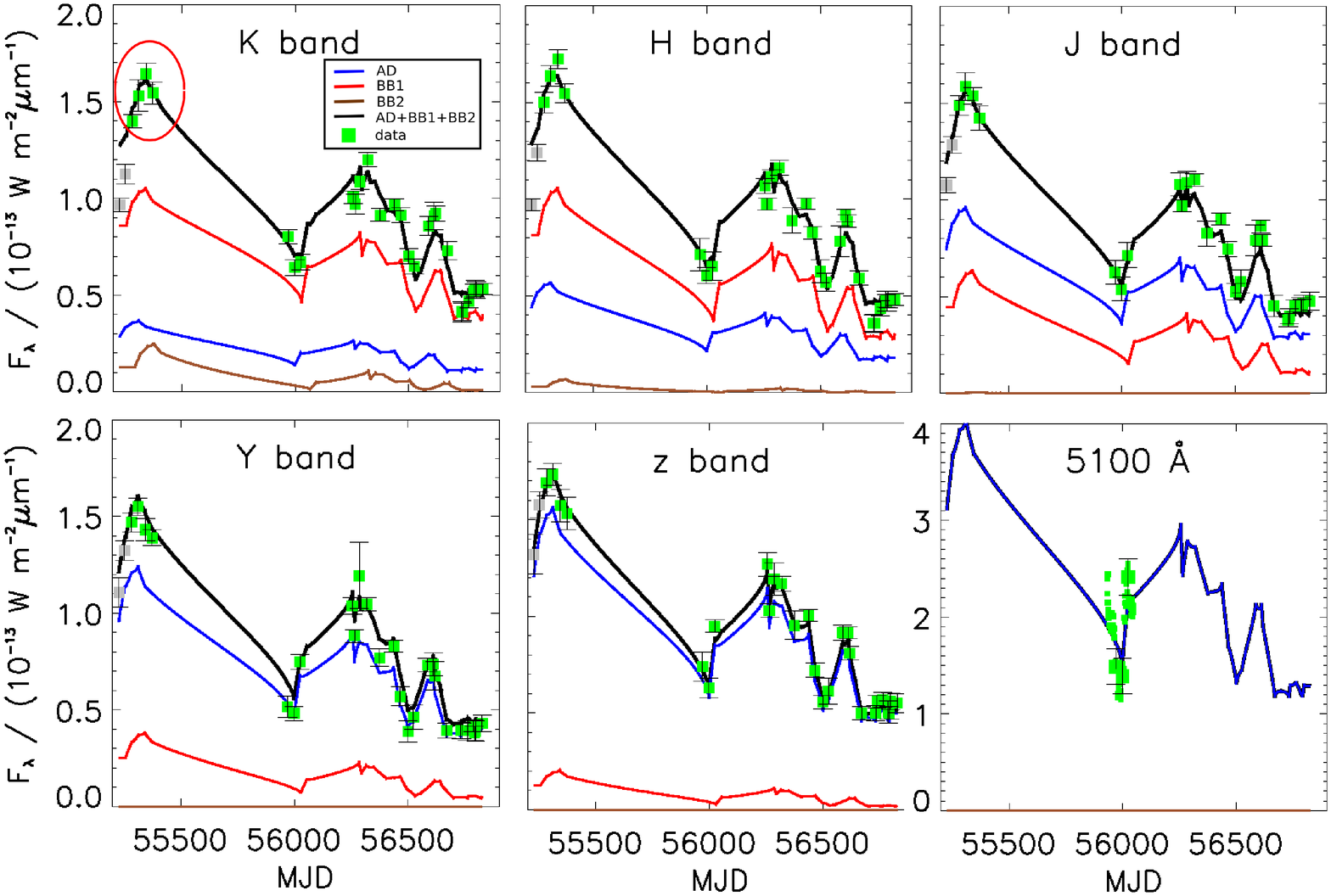}
   \caption{\label{fig:Fit_2BB} Results of our 2BB model fit. Plotted are the $K,H,J,Y,z$, and 5100 \AA\ data (green) over time. The bigger dots with error bars in the 5100 \AA\ panel denote the data points that coincide with our NIR epochs. The red line represents the contribution of the inner blackbody component (at 29 light-day distance) for each band, the brown line is the second blackbody (at 67 light-days), blue the AD contribution, and the black line is the sum of all three components. Epochs 1, 2, and 3 are not included in the evaluation of the fit (as $t < \tau_2$). The resulting 2BB model fit only differs significantly from the 1BB model fit in the high-flux epochs 4-6, in the $K$ band, as indicated by a red circle.}
    \end{figure*}

The results of our multiepoch, multiwavelength fit is shown in a temporal plot in Fig.~\ref{fig:Fit_1BB} for the 1BB model and in Fig.~\ref{fig:Fit_2BB} for the 2BB model. For the 1BB model, the best-fit lag is $\tau=31.0 \pm 1.6$ days, and the best-fit value for the power-law slope is $\alpha=1.63\pm0.04$ (cf.~Table \ref{tab:pars} for values of the other parameters), nicely matching the $F_{\nu}\propto \nu^{1/3}$ law expected from a standard Shakura-Sunyaev accretion disk. The initial blackbody temperature has a best-fit value of $T_{0}=1436 \pm 15$ K, and then evolves (by construction) as in Eq.~\ref{eq:T}. 
Epochs 1 and 2 are not included in the evaluation of the fit, as in our model, $T$ evolves only for $t \geq \tau$. We find $\chi^2_{red}=1.87$ for this fit, indicating that we have probably slightly underestimated our measurement errors or that the model is too simple.\\
For the 2BB model, we find best-fit values of $\tau_1=29.3 \pm 1.9$ days, $T_{0,1}=1479 \pm 27$ K, $\tau_2=67.2 \pm 8.3$ days, $T_{0,2}=698 \pm 64$ K, $\alpha=1.65 \pm 0.04$, and a reduced $\chi^2$ value of $\chi^2_{red}=1.89$. Here, epochs 1-3 are not included into the fit, because $T_2$ evolves only for $t \geq \tau_2$. The best-fit values for the complete set of parameters are given in Table \ref{tab:pars}. For both the 1BB and the 2BB model, the data can be fit well with a stable time lag.\\

In Fig.~\ref{fig:4epochs}, we show the results from our 1BB fit in the spectral domain, for the epochs 3, 5, 9, and 10. It is clear how the observed SED changes from epochs 3 to 5, and we can see a hot dust bump emerging, with the peak of the blackbody emission shifted to lower wavelengths ($\lambda\approx 1.9 \mu$m), thus indicating our detected temperature increase. Though degeneracies between the blackbody constant and the blackbody temperature are non-negligible (in the single-epoch fits, where the temperature can evolve freely, the correlation coefficient is as high as $\approx -0.75$), visual inspection of the data and 
the systematic changes in the NIR color, correlating with the delayed AD 
brightness, underline the actual temperature increase. The hot dust peak is shifted to much longer wavelengths ($\lambda\approx 2.5 \mu$m) in epoch 9, and until epoch 10, we again observe rising temperatures.\\

\begin{figure*}[htb]
   \centering
  \includegraphics[width=16.cm]{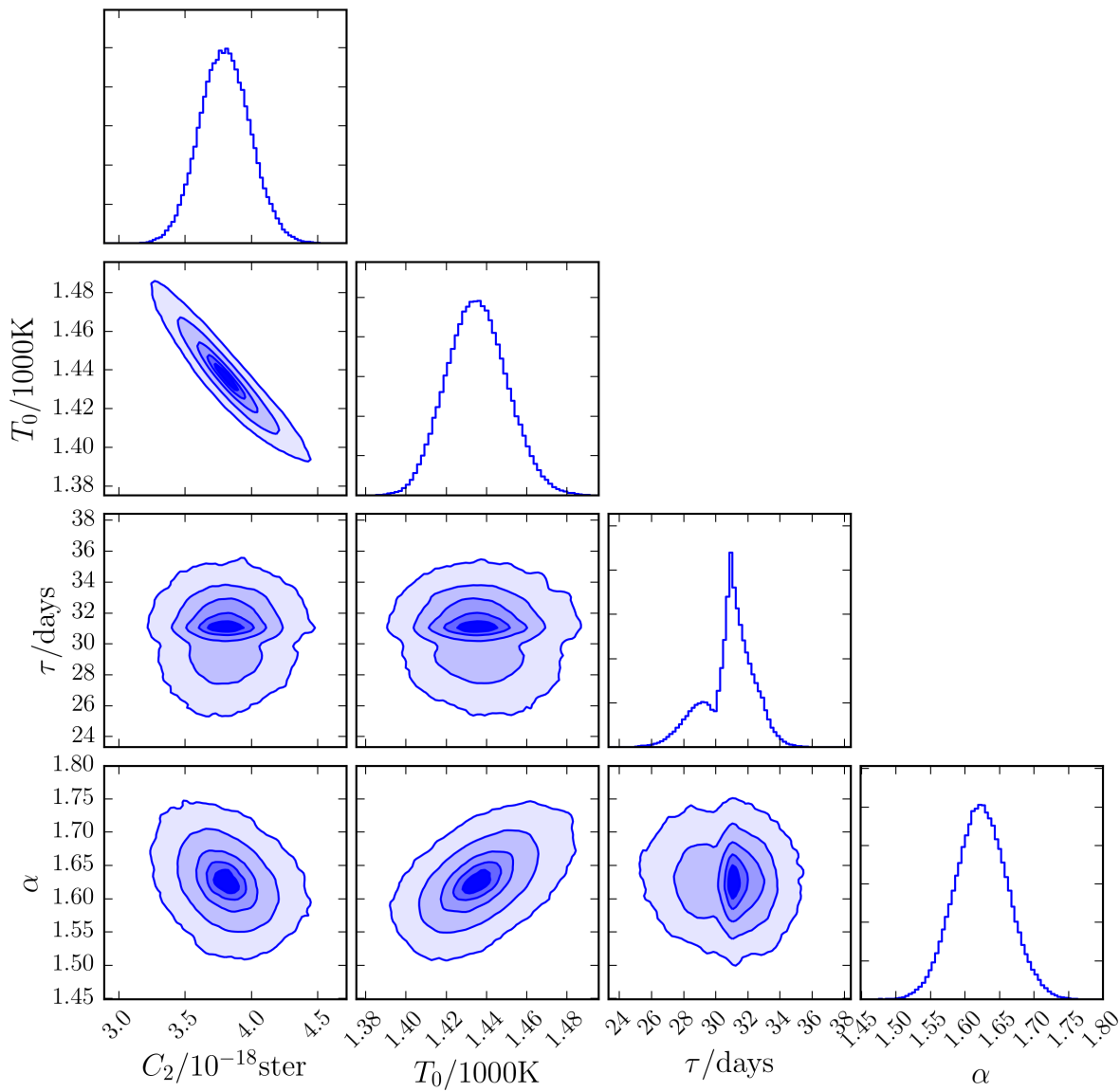}
   \caption{\label{fig:cornerplot_all} Marginalized posterior probability distributions for the four parameters $C_2, T_0,\tau$, and $\alpha$ of the 1BB model fit, with the 1-dimensional projections shown along the diagonal, and the 2-dimensional projections in the other panels. Contours mark the 10\%, 25\%, 50\%, 85\% and 99\% confidence intervals.}
    \end{figure*}

In Fig.~\ref{fig:cornerplot_all}, we show the marginalized posterior probability distributions for the parameters $C_2, T_0, \tau$, and $\alpha$.  As can be seen, $C_2$ and $T_0$ are strongly anti-correlated. However, any increase or decrease in $C_2$, hence decrease or increase in $T_0$, will only shift the resulting temperature curve in the vertical direction, owing to our approach Eq.~\ref{eq:T}. The significance of the temperature variations justifying this approach has already been shown with our single-epoch fits (Fig.~\ref{fig:LzAD_vs_T1}) and can also be seen from the SEDs shown in Fig.~\ref{fig:4epochs}. The reverberation delay $\tau$ does not seem to show significant correlations with any other parameter, while $\alpha$ is slightly anti-correlated  resp.~correlated with $C_2$ resp.~$T_0$. Interestingly, we observe a multimodal pdf of the reverberation lag $\tau$. This is on the one hand caused by slightly different reverberation delays in 2010 and 2013-2014 (see Sect.~\ref{subsec:VarLag}), but mainly by a strong bimodality in $\tau$ in the 2012-2014 part of the data set (epochs 7-29, see Fig.~\ref{fig:cornerplot_firstsecond}). Because that second period dominates the fit due to a higher amount of data points and lower photometric errors, the bimodal pdf is also visible in the fit of the complete data set. This bimodality is discussed in more detail in Sect.~\ref{subsec:VarLag}.

\subsection{Further structure function models and the time-variable power-law index}
\label{subsec:alpha_var}

As described in Sect.~\ref{subsec:interp}, we tested for the robustness of all our performed multiepoch, multiwavelength fits using three different structure function models. It turned out that the results are highly stable under the exchange of the structure function model or its particular parameters. In Table    \ref{tab:pars}, the results are only listed for the power-law model, while Table \ref{tab:pars_first9_second_struct} shows the influence of the structure function on a subset of our results.\\ 
We also tested the influence of the AD power-law slope $\alpha$ on our results. As an alternative approach to keeping $\alpha$ fixed over the whole time and flux range, we allowed for a varying $\alpha$. Since the continuum emission from the AD is found to get harder as the AD brightens (see e.g.~\citet{Trev2001} and references therein), we performed an alternative 1BB fit allowing for a varying power-law slope of the form $\alpha_{var}(t)=\alpha + \alpha_1\cdot L_z(t)/\langle L_z(t) \rangle$. We obtained best fit values $\alpha=1.56 \pm 0.07$ and $\alpha_1=0.10 \pm 0.07$, while all other parameters stayed nearly the same (Table \ref{tab:pars}). 
At this point, we note that unfortunately the power-law slope is only well-determined in those epochs where optical measurements were also available (epochs 7-9). Thus, our applied approach of fitting a varying $\alpha$ to our data might be insufficient to determine the true variability range of the AD power-law slope. Therefore, we alternatively used published empirical relations between $\alpha$ and the AD luminosity. According to \citet{Trev2001}, who analyzed multiepoch data of a sample of quasars, there is a relation $\Delta\alpha=a+b\Delta(\mbox{log}F_{\nu})$ between the change of the spectral index and the logarithmic optical continuum flux change of the AD, with $a=(-8.49\pm5.50)\cdot10^{-2}$ and $b=2.55\pm0.75$. Specifically for NGC 4151, \citet{Fan1984} report the relation $\alpha=b\cdot \mbox{log} F_{\nu}$ with $b\approx 4$. Making use of the inferred $\alpha$ values in epochs 7-9 and the flux differences of the $z$ band signal with respect to the $z$ flux in these three epochs, we applied the cited two models for deriving an alternative, more representative evolution of $\alpha$ over the whole time and flux range. Thus, instead of fitting the evolution of $\alpha$ to our data, which may problematic because of the absence of optical data in most epochs, we now use those two models as input for our fit. We find our results to be qualitatively robust under the use of these different models (also see Sect.~\ref{subsec:VarLag}), even though values as high as $\alpha\approx 3.3$ are reached in the high-flux epochs. The influence of the particular choice of $\alpha$ for a subset of our results is shown in Table \ref{tab:pars_first9_second_alpha}.

\section{Discussion}
\label{sec:Discussion}
\subsection{Single-blackbody model}

As for our 2010 data of NGC 4151, we observe a 
significant rise of the emission and temperature of the innermost hot dust, following states of increased AD brightness. The hot dust temperature follows the AD flux with a time delay of roughly 31 days.
This indicates that the  hot dust in NGC 4151 currently observed is simply heated up
by increased AD irradiation and not destroyed owing to sublimation. In the case of significant dust destruction, one would expect an increase in the reverberation delay, which is not seen in our data. Obviously, the major part of the hot dust in the nucleus of NGC 4151 is not located at its current sublimation radius, but is cooler than sublimation temperature.\\
There are strong indications that the hot circumnuclear dust around AGNs mainly consists of large graphite grains ($\approx 0.2 \mu$m grain size), with sublimation temperatures $\gtrsim$ 1500K \citep{Gask2004, Kish2007, Kish2011a}. Since our inferred dust temperatures do not reach 1500K (see Fig.~\ref{fig:LzAD_vs_T1} and Fig.~\ref{fig:T1multi_vs_T1single}), it is to be expected that we see no dust sublimation in our data.\\

Limitations of our method are given by our $z$-band sampling, and thus the interpolation of our input AD signal. Because, for example, the minimum observed in $JHK$ around MJD 56700 is missed in the $z$-band observations (therefore in the interpolated AD signal), the resulting model fits the $JHK$ fluxes around MJD 56700 very poorly. Apart from this mismatch, the 1BB model already fits the data remarkably well within the errors. Only in the epochs 4-6 are the modeled $K$-band fluxes systematically lower than the actually measured $K$-band peak fluxes. 

\subsection{Two-blackbody model}
\label{Discussion_2BB}
In the 2BB model, epochs 4-6 are fit better, indicating that a second BB component might actually contribute significantly to the observed $K$-band flux. However, we must point out that the 
goodness of the fit is not improved when using the 2BB model ($\chi^2_{red}$=1.89) instead of the
1BB model ($\chi^2_{red}$=1.87)). This is not surprising, since the weight of the few 2010 $HK$ data points 
is low compared to the global data set.\\
In our Paper I, we already argued that the hot, NIR dust around NGC 4151 might 
be substantially radially extended and better represented by more than one blackbody component.\\

However, when including data at longer wavelengths, i.e.~the WISE W1-W3 bands, we must see
that our 2BB model has to be rejected\footnote{The WISE data were not included in the $\chi^2$ of the fit, because the aim was to test for an extended structure in the NIR regime up to the $K$ band. Nevertheless, any resulting 2BB model would have to match the WISE fluxes as well.}. From Figs.~\ref{fig:WISE_1BB} and \ref{fig:WISE_2BB}, it becomes apparent that the fluxes are fit better with the 1BB model.\\

This finding agrees with the data of \citet{Bur2009}. The unresolved point source that they measure with $N$-band interferometry and which they attribute to the inner rim of the hot dust torus (at 0.04 - 0.05 pc $\approx$ 50 light-days) is apparently the same source that constitutes our observed NIR fluxes, because the observed Burtscher PS fluxes are consistent with the SED of our 1BB model (Fig.~\ref{fig:WISE_1BB}). 
In Figs.~\ref{fig:WISE_1BB} and \ref{fig:WISE_2BB}, we went on to overplot the fluxes of an extended Gaussian source measured by \citet{Bur2009}, and the sum of both components (PS + Gaussian). The extended source is interpreted as the warm component of the clumpy torus located farther out, at 2.0 $\pm$ 0.4 pc $\approx$ 240 light-days. Because this extended source is not resolved in the WISE photometry, it contributes to the WISE PSF flux of the W3 band (which matches the Burtscher total flux at 12.5 $\mu$m), while the W1 and W2 bands obviously show no significant warm dust contribution. A clear excess of the observed PS + Gaussian fluxes over the modeled 1BB fluxes can only be seen from the $N$ band ($\lambda\geq 8\mu$m) on. Here, the single blackbody approximation is no longer sufficient, but a second blackbody component contributes to, or even dominates, the measured fluxes. This is, however, not the $698 \pm 64$K blackbody component from our 2BB fit, but a blackbody of lower temperature ($T=285^{+25}_{-50}$) that is located farther out \citep{Bur2009}.
A second blackbody component within a 100 light-day distance from the source (which we set as an upper limit in the prior for $\tau_2$, see caption of Table \ref{tab:priors}) is thereby clearly ruled out. \\


\subsection{Variable time lag}
\label{subsec:VarLag}

\begin{figure*}[htb]
   \centering
  \includegraphics[width=16.cm]{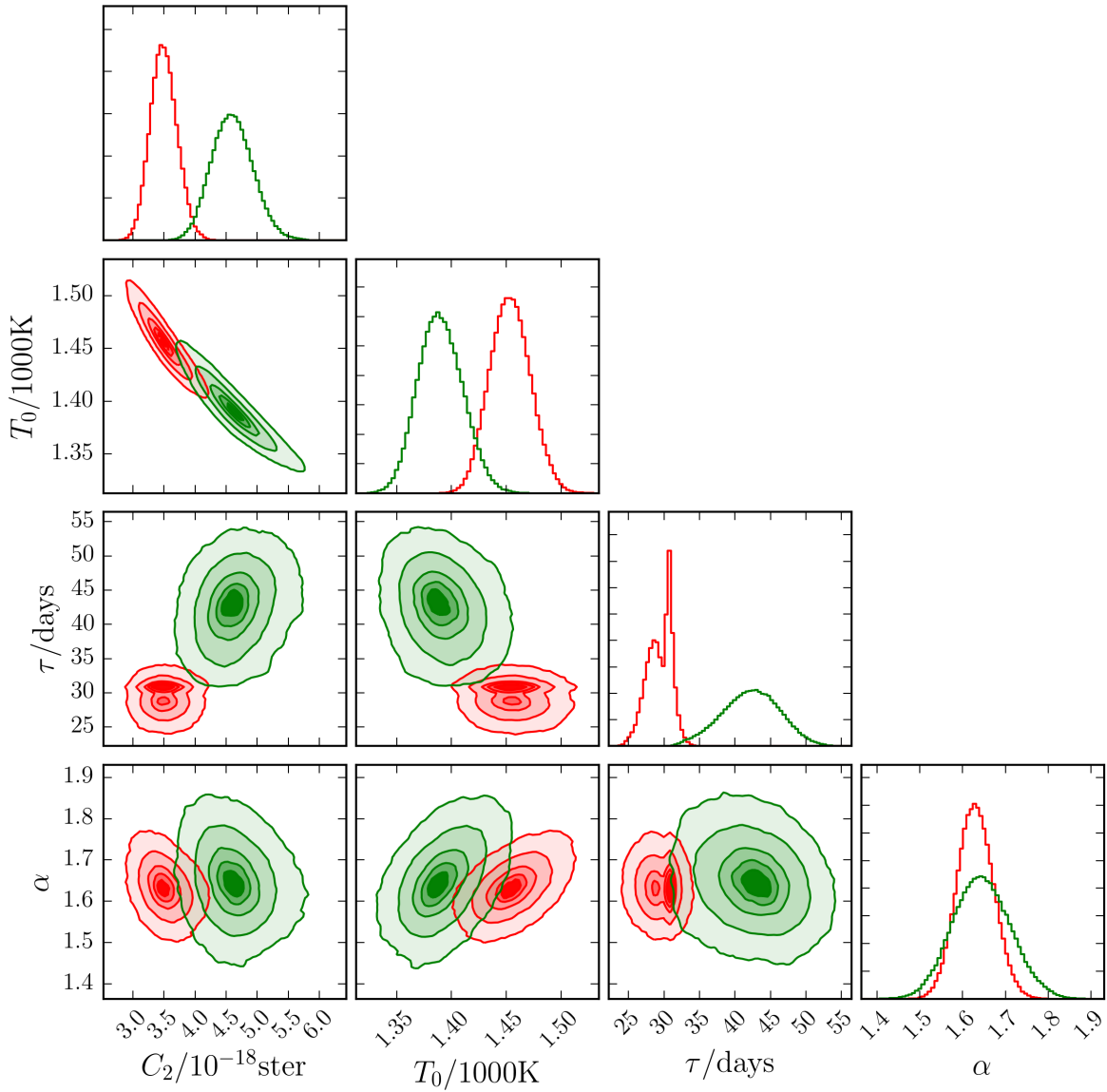}
   \caption{\label{fig:cornerplot_firstsecond} Marginalized posterior probability distributions for the four parameters $C_2, T_0,\tau$, and $\alpha$ of the 1BB model fit, for the first part of the data set (epochs 1-9) in green and the second part (epochs 7-29) in red. The 1-dimensional projections are shown along the diagonal, and the 2-dimensional projections in the other panels. }
    \end{figure*} 

\begin{table*}
\caption{Global mean and errors for the parameters in the first period (epochs 1-9) and in the second period (epochs 7-29) for the 1BB model. The errors are given by $\Delta x=\sqrt{\hat{V}}$ for each parameter. In particular, we show the influence of different structure function models discussed in Sect.~\ref{subsec:interp} on our results. Here, DRW1 refers to the damped random walk model with $\sigma$=0.15 and $\tau$=230, DRW2 to the damped random walk model with $\sigma$=0.03 and $\tau$=10. All results are shown for constant $\alpha$.
}       
\label{tab:pars_first9_second_struct}      
\centering          
\begin{tabular}{l | c | c | c | c | c | c | c }     
\hline       
\hspace{-0.15cm}Parameter & Unit & ep.~1-9 & ep.~1-9 & ep.~1-9  & ep.~7-29 & ep.~7-29 & ep.~7-29\\
 & & power law & DRW1  & DRW2  & power law & DRW1  & DRW2 \\
\hline              
   \hspace{-0.15cm}$C_1$ & -- & 0.78 $\pm$ 0.01 & 0.78 $\pm$ 0.01 & 0.78 $\pm$ 0.01 & 0.78 $\pm$ 0.01 & 0.78 $\pm$ 0.01& 0.78 $\pm$ 0.01\\ 
\hspace{-0.15cm}$\alpha$ & -- & 1.65 $\pm$ 0.07& 1.65 $\pm$  0.08& 1.65 $\pm$ 0.08& 1.63 $\pm$ 0.04 & 1.64 $\pm$ 0.06& 1.64 $\pm$ 0.05\\ 
\hspace{-0.15cm}$C_2$ & $10^{-18} \mbox{ster}$ & 4.63 $\pm$ 0.33 & 4.66 $\pm$ 0.35& 4.67 $\pm$ 0.36& 3.52 $\pm$ 0.22 & 3.55 $\pm$ 0.34& 3.55 $\pm$ 0.32\\ 
\hspace{-0.15cm}$T_{0,1}$ & $1000 \mbox{K}$ & 1.392 $\pm$ 0.020& 1.393 $\pm$ 0.027 & 1.389 $\pm$ 0.026  & 1.455 $\pm$ 0.018&  1.449 $\pm$ 0.017 & 1.443 $\pm$ 0.019\\ 
\hspace{-0.15cm}$\tau_1$ & $\mbox{days}$ & 42.5 $\pm$ 4.0 & 42.4 $\pm$  4.6 & 42.2 $\pm$ 4.4 & 29.6  $\pm$ 1.7 & 30.1 $\pm$ 2.1 & 30.2 $\pm$ 2.0\\ 
\hspace{-0.15cm}$v_1$ & -- & 0.96 $\pm$ 0.04 & 0.91 $\pm$ 0.03 & 0.91 $\pm$ 0.04& 0.81 $\pm$ 0.03 & 0.77 $\pm$ 0.03& 0.76 $\pm$ 0.03\\
\hline        
\end{tabular}
\end{table*}

\begin{table*}
\caption{Global mean and errors for the parameters in the first period (epochs 1-9) and in the second period (epochs 7-29) for the 1BB model. The errors are given by $\Delta x=\sqrt{\hat{V}}$ for each parameter. In particular, we show the influence of the evolution of $\alpha$ (see Sect.~\ref{sec:Results}) on the robustness of our results. All results are shown for the power-law structure function model.
}       
\label{tab:pars_first9_second_alpha}      
\centering          
\begin{tabular}{l | c | c | c | c | c | c | c }     
\hline       
\hspace{-0.15cm}Parameter & Unit & ep.~1-9 & ep.~1-9 & ep.~1-9  & ep.~7-29 & ep.~7-29 & ep.~7-29\\
 & & $\alpha=$const & $\alpha$ (Trev.~2001) & $\alpha$ (Fan.~1984) & $\alpha=$const & $\alpha$ (Trev.~2001) & $\alpha$ (Fan.~1984)\\
\hline              
   \hspace{-0.15cm}$C_1$ & -- & 0.78 $\pm$ 0.01 & 0.75 $\pm$ 0.01 & 0.70 $\pm$ 0.01  & 0.78 $\pm$ 0.01 & 0.79 $\pm$ 0.01 & 0.77 $\pm$ 0.01\\ 
\hspace{-0.15cm}$\alpha$ & -- & 1.65 $\pm$ 0.07& --  & -- & 1.63 $\pm$ 0.04 & -- & --\\ 
\hspace{-0.15cm}$C_2$ & $10^{-18} \mbox{ster}$ & 4.63 $\pm$ 0.33 & 4.83 $\pm$ 0.43 & 4.42 $\pm$ 0.28 & 3.52 $\pm$ 0.22 & 3.80 $\pm$ 0.35 & 3.70 $\pm$ 0.21\\ 
\hspace{-0.15cm}$T_{0,1}$ & $1000 \mbox{K}$ & 1.392 $\pm$ 0.020& 1.405 $\pm$  0.036 & 1.451 $\pm$ 0.018  & 1.455 $\pm$ 0.018 &  1.463 $\pm$ 0.027 & 1.505 $\pm$ 0.039\\ 
\hspace{-0.15cm}$\tau_1$ & $\mbox{days}$ & 42.5 $\pm$ 4.0 & 36.5 $\pm$ 5.3 & 33.9 $\pm$ 4.2 & 29.6  $\pm$ 1.7 & 28.3 $\pm$ 2.8 & 25.9 $\pm$ 2.5\\ 
\hspace{-0.15cm}$v_1$ & -- & 0.96 $\pm$ 0.04 & 1.09 $\pm$ 0.04 & 1.15 $\pm$ 0.05 & 0.81 $\pm$ 0.03& 1.03 $\pm$ 0.02 & 1.24 $\pm$ 0.04 \\
\hline        
\end{tabular}
\end{table*}

\begin{figure*}[htb]
   \centering
  \includegraphics[width=17.cm]{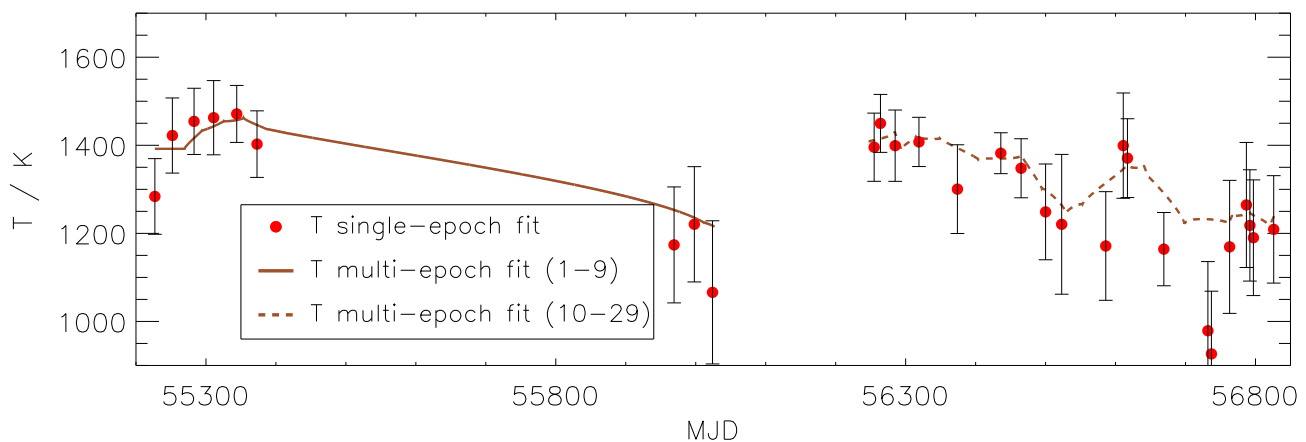}
   \caption{\label{fig:T1multi_vs_T1single} Hot dust temperature $T_{single}$ derived from the single-epoch decomposition, where $T$ is a free parameter in each epoch, versus the temperature $T_{multi}$ from the multiepoch fits, in which $T$ evolves according to Eq.~\ref{eq:T} and only the initial temperature $T_0$ is a free parameter. For epochs 1-9, the temperature resulting from the epoch 1-9 fit is plotted, for epochs 10-29 the temperature  from  the epoch 7-29 fit. The minimum that is seen in $T_{single}$ around MJD 56700 (and also in the corresponding $JHK$ flux minima) is not covered within the sampling of the $z$-band data (see Figs.~\ref{fig:Fit_1BB} and \ref{fig:fluxes}), therefore we observe a substantial discrepancy between $T_{single}$ and $T_{multi}$ around that date.}
    \end{figure*}

Besides a second blackbody component contributing to the NIR fluxes,
there might be another explanation for the missed 2010 $K$-band peak by our fit.
The resulting lag of $\tau_1=31.0 \pm 1.6$ days found for the 1BB model fits our global data very well. This holds for all data except for the $K$-band peak in epochs 4-6. Here, it looks as if the actual lag might still be a bit higher. If we split the data set into two subsets (set 1: epoch 1-9 (2010 - beginning of 2012 data), set 2: epochs 7-29 (2012 - 2014 data)\footnote{We split whole data set into these two overlapping data sets because, while it was obvious that the delay in epochs 1-6 would be longer than the delay in epochs 10-29, a priori the delay of epochs 7-9 was not apparent, because fitting only 3 epochs is rather ambiguous.}), 
we find 
that the fitted hot dust reverberation delay is $\tau^{2010}=42.5 \pm 4.0$ days if we include only epochs 1-9 (see Table \ref{tab:pars} for the other parameter values), consistent with \citet{Hoe2011} and \citet{Kish2011b}. For the epoch 7-29 fit, we get a best-fit delay of $\tau^{2013}=29.6 \pm 1.7$ days (also see Table \ref{tab:pars}). We excluded that this decrease in $\tau$ is merely an effect of the improved sampling in the second period. The blackbody constant also changes between the two different fits, from $C_2^{2010}=4.63 \pm 0.33$ to $C_2^{2013}=3.52 \pm 0.22$, and the initial temperature from $T_{0}^{2010}=1392 \pm 20 \mbox{K}$ to $T_{0}^{2013}=1455 \pm 18 \mbox{K}$.
However, $T_0^{2013}$ refers to the initial temperature in epoch 1 and thus has no real physical meaning for the second-period fit, since only epochs 7-29 are included. Figure \ref{fig:T1multi_vs_T1single} shows the resulting temperature evolution of this fit.
 Interestingly, while all other parameters stay nearly the same, the variability factor also changes between the two fits, from $v^{2010}=0.96 \pm 0.04$ to $v^{2013}=0.81 \pm 0.03$, which might indicate that the 
AD illumination is less efficiently reprocessed by the hot dust in the second period.\\

As discussed in Sect.~\ref{sec:Results}, we tested the influence of different structure-function models and different evolutions of the AD power-law index $\alpha$ on our results. It seems particularly important to test the robustness of the apparent decrease in the reverberation lag under the different models. While the parameter values are very insensitive to the different structure function and $\alpha$ models for the global 1BB model, which includes all epochs, the parameters inferred for the epoch 1-9 fit and epoch 7-29 fit do undergo certain changes when applying the different structure functions and $\alpha$ models. The influence of the various structure function models on our results are shown in Table \ref{tab:pars_first9_second_struct}, while the influence of $\alpha$ is shown in Table \ref{tab:pars_first9_second_alpha}. We do see a change in the derived absolute parameter values, especially the time lag seems to be sensitive to the applied variability model of $\alpha$. Nevertheless, relative to each other, the inferred lags for the epoch 1-9 and epoch 7-29 fits and the decrease in the delay remain unchanged. While for a constant $\alpha$ (and different structure functions), the lag decreases from $\tau^{2010}\approx 43$ days to $\tau^{2013}\approx 30$ days, we infer $\tau^{2010}\approx 37$ days, $\tau^{2013}\approx 28$ days using the model according to \citet{Trev2001}, and $\tau^{2010}\approx 34$ days, $\tau^{2013}\approx 26$ using the one presented by \citet{Fan1984}. Thus,  our results are qualitatively unaltered: we see a significant decrease in the reverberation delay from 2010 to 2013-2014.\\
 
\begin{figure}[htb]
   \centering
  \includegraphics[width=9cm]{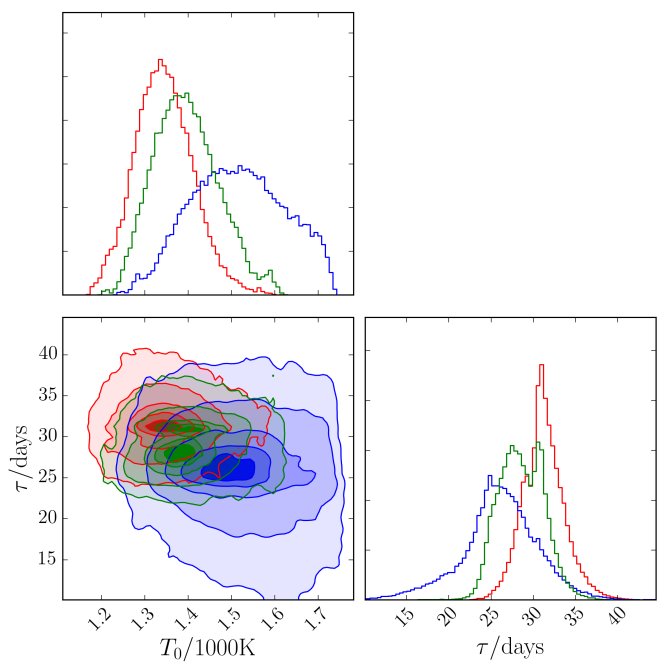}
   \caption{\label{fig:pdfs_JHK} Marginalized posterior probability distributions for the parameters $T_0$ and $\tau$ of the epoch 7-29 1BB model fit, obtained by fitting the $J$, $H$, and $K$ bands separately (also see text of this section). The $J$ band fit is shown in blue, $H$ band in green, and $K$ band in red. The 1-dimensional projections of the pdfs are shown along the diagonal, and the 2-dimensional projections in the other panel.}
    \end{figure} 

 In Fig.~\ref{fig:LzAD_vs_T1}, the temperature evolution resulting from the epoch 1-9 and epoch 7-29 fits is plotted versus the temperature from the single-epoch decomposition (Eq.~\ref{eq:Model_single}). Naively one would expect that due to the decreased AD radiation by 50\% on average from 2010 to end of 2012-2014 ($\langle L_z\rangle^{2013} / \langle L_z\rangle^{2010} \approx 0.49$), the blackbody temperature would have decreased by a factor $0.5^{1/4}\approx 0.84$ following $L\propto T^4$. This temperature decrease due to the dimmer AD would be roughly balanced by a temperature increase due to a new dust location farther inside by a factor of $\sqrt{\tau^{2010}/\tau^{2013}}\approx1.18$. Thus, one would expect the blackbody temperature for 2010 to be on the same average level as at the end of 2012-2014. Nevertheless, we observe a decreased average temperature in the second part of the data set ($\langle T\rangle^{2013} / \langle T\rangle^{2010} \approx 0.86$) in the single-epoch fits. This is confirmed by the temperature curve of our multiepoch fit for epochs 7-29 and achieved through a lower variability factor with $v^{2013}/v^{2010}\approx0.84$. Obviously, in the second part of the data set, the dust is heated less efficiently by the AD radiation estimated from the $z$ band flux. The physical reasons for this could be an increase in dust grain size or a geometrical cause. In a model proposed by \citet{Czer2011}, the dust clouds resulting from an AD wind only become exposed to the AD radiation in the process of moving farther outside, because dust that is still located farther inside tends to rise only a short distance above the disk, whereas the dust farther outside has a greater height. The innermost dust would then be heated less efficiently assuming an anisotropic radiation characteristic of the AD. The discussed parameter changes between the two periods might point to a changed dust distribution.\\

In Fig.~\ref{fig:cornerplot_firstsecond}, we show the marginalized posterior probability distributions of the parameters $C_2, T_0, \tau$, and $\alpha$ for the epoch 1-9 fit and the epoch 7-29 fit . Clearly, we see a significantly decreased reverberation lag from $\tau^{2010}=42.5 \pm 4.0$ in 2010 to $\tau^{2013}=29.6 \pm 1.7$ in 2012-2014. As already mentioned in Sect.~\ref{sec:Results}, the pdf of $\tau^{2013}$ is bimodal. Interestingly, no bimodality is observed for $\tau^{2010}$. Here, the pdf seems perfectly unimodal. Possibly, the bimodality is washed out by the higher photometric errors in the first period. Increasing the errors in the second period to the level of the epoch 1-9 errors, however, only slightly washes out the bimodality but does not make it vanish completely. We furthermore excluded that the detected bimodality in epochs 1-9 in contrast to the missing bimodality in epochs 7-29 is merely due to the improved sampling in the second period. This supports our supposition that we see a changed dust distribution in the second period.\\ 


The resolved bimodality in $\tau^{2013}$ is caused by slightly different delays for the bands $JHK$ (constituting the blackbody emission), as confirmed by performing separate fits for each band\footnote{We fixed $C_1, \alpha$ and $v$ to the best fit values of the epoch 7-29 fit for these single-band fits, to avoid high degeneracies between the parameters.}, i.e.~$\tau_J^{2013}=26.1 \pm 4.8$, $\tau_H^{2013}=28.9 \pm 2.6$ and $\tau_K^{2013}=31.4 \pm 2.4$, pointing to a slight radial extent of the dust within a narrow region around $\tau^{2013}$. Interestingly, what we observe is not just a smoothly extended structure, but at least two separate blackbodies at two discrete, nearby, but different radii. This feature is visible in Fig.~\ref{fig:pdfs_JHK}, especially in the $H$ band. While the two-dimensional projections are hard to disentangle by the eye, in the right panel of Fig.~\ref{fig:pdfs_JHK} one can nicely see an "inflection point" of the two different reverberation delays around $\tau\approx 30$ days.  The pdf peaks at two different delays, $\tau_{inner}\approx 26$ days and $\tau_{outer}\approx 31$ days. In the $K$ band, $\tau_{outer}$ dominates the pdf, while $\tau_{inner}$ is slightly visible as well. In the $J$ band, the pdf is clearly dominated by the shorter lag, while the longer lag can only be marginally "resolved" due to a slight asymmetry in the pdf. The intersection point between the pdfs of the two different delays always occurs at $\tau\approx 30$ days, for all three bands, and $H$ and $K$ even exhibit a minimum there. As expected, the fitted blackbody temperatures of the three bands decrease with wavelength, from $T_{0,J}^{2013}=1521 \pm 105$K, $T_{0,H}^{2013}=1392 \pm 74$K, and $T_{0,K}^{2013}=1351 \pm 102$K. Due to missing information on the color in the single-band fits, the degeneracies between $T_0$ and $C_2$ are extremely high, so the fitted temperatures should not be taken too seriously. The detected 2BB structure within a narrow range around $\tau^{2013}$ does not contradict the result of our previous 2BB model fit, which was mainly performed to test for a significantly radially extended dust distribution rather than a compact distribution. The two distinct blackbodies resolved within the epoch 7-29 fit still represent a fairly compact dust distribution, since they are very close, and are therefore not in conflict with our finding from the 2BB model fit.\\

Our results seem to indicate a decreased time delay from 2010 (MJD 55300-55400) to 2012-2013 (MJD 56000-56300). This stands in strong contrast to interferometric observations by \citet{Kish2013}, who measured an increased delay, from roughly 40-50 days in 2010 to 70-80 days in 2012 -- a delay that is clearly not consistent with our data.\\ 
It is possible that this apparent shift in the time lag only represents the
complicated and dynamic dust morphology with the dust being clearly extended and with clouds moving turbulently within this dust distribution. It seems possible that even slight changes in the dust distribution through motion of single clouds could cause noticeable changes in the measured torus response, e.g.~by means of shielding.
However, the shift in time lag could also mean that the dust radius has indeed decreased by inflow of dust from outside due to the low luminosity state of the AD and thereby decreased dust sublimation radius. Alternatively, BEL clouds could have been launched from farther inside due to decreased AD luminosity. Dust condensation only occurs when the clouds have expanded to roughly three times their initial radius, which happens at a distance $d\approx1000\cdot d_0$, with $d_0$ the initial distance of the clouds from the source \citep{Elv2002}.
Dust condensation would set in after roughly three to nine years from the initial launching of the clouds.
Interestingly, our measured decrease in time lag seems to continue the decline from $\tau\approx 60$ days observed end of 2008 \citep{Pot2010} to $\tau\approx 50-45$ days in 2009/2010 \citep{Kish2009, Kish2011b}. This total decrease in $\tau$ from 2008-2012 seems to track a decline in the AD flux from 2003 until 2006 with a delay of roughly five to six years (see Fig.~3 in \citet{Kish2013}, thus perfectly matching the timescale expected within the accretion wind scenario.\\
An upper limit for the infall velocity of dust produced by stars, which is moving in from outside, is the free fall velocity $v_{free}=\sqrt{2GM/R_{dust}}$ 
onto the central black hole ($M_{BH}\approx 4.6\cdot10^{7}M_{Sun}$, \citealt{Ben2006}). We obtain an infall velocity of $v_{free}\approx 0.01 c$ at the location of the hot dust torus. It would thus take $\gtrsim$ 1300 days to decrease the dust radius from 43 to 30 days. This time span seems only marginally possible within our estimate. Although these qualitative arguments seem to favor the accretion wind scenario over the inflow model, we note that detailed modeling of the different scenarios is needed to reliably conclude on the dust formation mechanism in AGNs.\\
Our measured 2010-2012 and 2012-2014 dust radii are inferred with one and the same method; however, the decrease over time of these measured radii will not fit the generally expected $R_{dust}\propto\sqrt{L_{AD}}$ relation, because it cannot be reasonably related to a decay in the momentary AD brightness delayed by $\tau$ (as also visible in Fig.~3 in \citet{Kish2013} for other observations of NGC 4151). A variable time lag (that possibly traces the averaged AD signal of several years before, but not the momentary AD signal) might help explain the intrinsic scatter found in the radius-luminosity relation of dust around AGNs.

\begin{figure}[htb]
   \centering
  \includegraphics[width=9.cm]{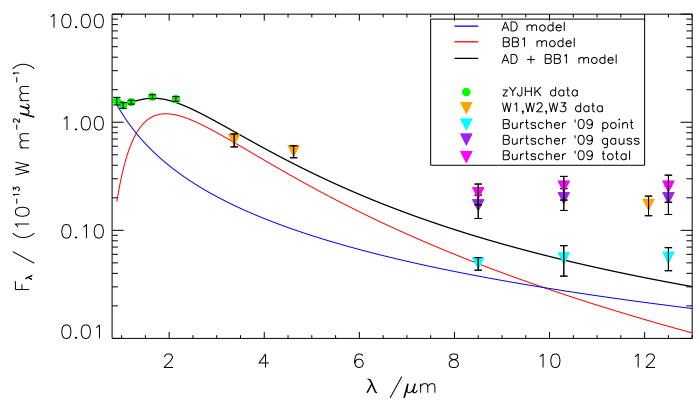}
   \caption{\label{fig:WISE_1BB} Resulting fluxes from the 1BB model fit in the high temperature epoch 5, plotted over wavelength. Overplotted are our derived NIR and MIR fluxes ($zYJHK$ and W1, W2, W3). The 1BB model roughly matches the MIR point source fluxes, given by \citet{Bur2009}. Since the model was fit to our $zYJHK$ data alone (also see Sect.~\ref{Discussion_2BB}), these are marked with green circles, while the other data are represented by triangles.}
    \end{figure}  

\begin{figure}[htb]
   \centering
  \includegraphics[width=9.cm]{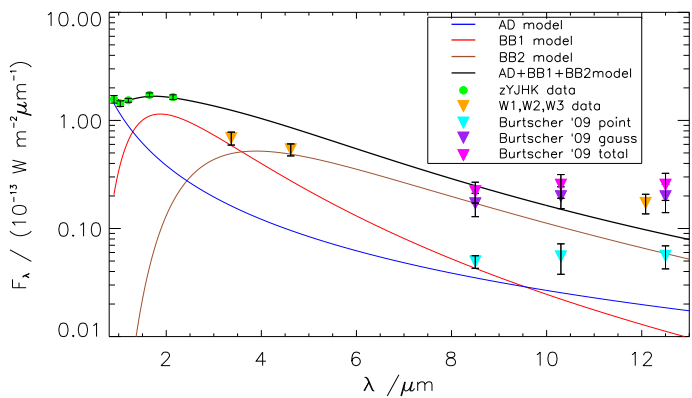}
   \caption{\label{fig:WISE_2BB} Resulting fluxes from the 2BB model fit in the high temperature epoch 5, plotted over wavelength. Overplotted are our derived NIR and MIR fluxes ($zYJHK$ and W1, W2, W3). While matching the observed $HK$ fluxes better than the 1BB model, the 2BB model total fluxes by far exceed the WISE W1, W2 measurements. A second blackbody component inside of 100 days can be clearly rejected. Again, the model was only fit to our $zYJHK$ data (marked with green circles, while the other data are represented by triangles).}
    \end{figure} 

\section{Conclusions}
\label{sec:Conclusions}
We presented updated results from our 2010-2014 NGC 4151 photometric and spectroscopic data, which are part of our AGN hot dust reverberation project to monitor the evolution of the hot dust temperature and reverberation lag around AGNs. Our findings for NGC 4151 are:
\begin{itemize}
\item Dust sublimation: If dust sublimation occurred in response to increased AD flux, it would happen at the inner edge of the dust distribution, thereby increasing the time delay. Although the AD brightness increased substantially within the observed time range, we see no signatures of any dust sublimation traced by our data. In particular, after detected AD flux increases by more than 30\% in 2010, and by roughly 100 \% from March 2012 to November 2012, we observed no increase in the hot dust reverberation delay, thus ruling out significant dust destruction for the time range 2010-2014. It seems that the hot dust in this galaxy is currently located beyond its sublimation radius.
\item Dust temperature: The hot dust emission, and moreover, the host dust temperature in NGC 4151 closely tracks the AD flux variations on a short-term response timescale, which is roughly one month on average. We measured maximum hot dust temperatures lower than 1500 K throughout 2010-2014. The large graphite dust grains that are typically assumed for the hot circumnuclear dust around AGNs have a sublimation temperature of $T_{sub}\gtrsim 1500K$. It thus seems perfectly consistent that we saw no dust sublimation in the observed period.
\item Dust distribution radius: On a long-term timescale, we measured a change of the hot dust reverberation delay of $\approx$13 days in two years. Detailed comparison of our results with seemingly comparable interferometric experiments reveals different variations in the derived dust radii. These apparent observational inconsistencies could imply that the real dust distribution in NGC 4151 is fairly complex, so that the comparison of hot dust radii measured with interferometry and reverberation is not straightforward.
\item Dust distribution: In our 2010 data, we saw slight indications of a second blackbody component of $\approx700$K, which could not be confirmed without broader wavelength coverage. Here, we presented a new analysis, now including WISE photometry, which rules out significant thermal radiation at 700K from radii smaller than 100 days.
\end{itemize}

To sum up, we found a decreased reverberation radius of the hot, circumnuclear dust in NGC 4151. Dust destruction in the observed epoch seems highly improbable from our data, but a slight change of the dust morphology seems likely. While the observed decrease in the reverberation delay matches the timescale expected within the accretion wind scenario perfectly, a radius decrease due to the inward motion of dust from outside seems only marginally possible, as estimated from an upper limit for the infall velocity of dust produced by stars (see Sect.~\ref{subsec:VarLag}). 
From our analysis, new dust formation in a cooling BLR wind appears to be more likely than a radius decrease due to inflow. We emphasize, however, that detailed modeling is indispensable to reliably distinguish the different dust formation scenarios in AGNs.
In our AGN host dust reverberation project, we monitor
roughly 25 additional bright and variable Seyfert 1 AGNs with the GROND
camera ($grizJHK$ bands, ESO La Silla) and in the optical with the
All-Sky Automated Survey for
Supernovae \citep{Shap2014}. It will be interesting to see whether our results for NGC 4151 apply to other AGNs as well.
\begin{acknowledgements} We thank the anonymous referee for valuable comments and suggestions that helped improve this manuscript. We are very thankful to R.~Andrae and M.~Fouesneau for helpful discussions and comments on this work.
We thank R.~van Boekel for providing a database of
main sequence star
atmospheres. K.~S.~ acknowledges support by ``IMPRS for Astronomy \& Cosmic
Physics
at the University of Heidelberg''. B.~M.~P.~and G.~D.~R.~are grateful for the support of the
US NSF through grant AST-1008882. B.S. is a Hubble, Carnegie-Princeton Fellow, and is supported by NASA through Hubble Fellowship grant HF-51348.001
awarded by the Space Telescope Science Institute, which is operated by the
Association of Universities for Research in Astronomy, Inc., for NASA,
under contract NAS 5-26555.
\end{acknowledgements}

%


\begin{thebibliography}{43}
\expandafter\ifx\csname natexlab\endcsname\relax\def\natexlab#1{#1}\fi

\bibitem[{{Alard}(2000)}]{Alard2000}
{Alard}, C. 2000, \aaps, 144, 363

\bibitem[{{Alard} \& {Lupton}(1998)}]{Alard1998}
{Alard}, C. \& {Lupton}, R.~H. 1998, \apj, 503, 325

\bibitem[{{Antonucci}(1993)}]{Ant1993}
{Antonucci}, R. 1993, \araa, 31, 473

\bibitem[{{Barvainis}(1987)}]{Bar1987}
{Barvainis}, R. 1987, \apj, 320, 537

\bibitem[{{Bauer} {et~al.}(2009){Bauer}, {Baltay}, {Coppi}, {Ellman}, {Jerke},
  {Rabinowitz}, \& {Scalzo}}]{Bau2009}
{Bauer}, A., {Baltay}, C., {Coppi}, P., {et~al.} 2009, \apj, 696, 1241

\bibitem[{{Bentz} {et~al.}(2006){Bentz}, {Denney}, {Cackett}, {Dietrich},
  {Fogel}, {Ghosh}, {Horne}, {Kuehn}, {Minezaki}, {Onken}, {Peterson}, {Pogge},
  {Pronik}, {Richstone}, {Sergeev}, {Vestergaard}, {Walker}, \&
  {Yoshii}}]{Ben2006}
{Bentz}, M.~C., {Denney}, K.~D., {Cackett}, E.~M., {et~al.} 2006, \apj, 651,
  775

\bibitem[{{Bentz} {et~al.}(2013){Bentz}, {Denney}, {Grier}, {Barth},
  {Peterson}, {Vestergaard}, {Bennert}, {Canalizo}, {De Rosa}, {Filippenko},
  {Gates}, {Greene}, {Li}, {Malkan}, {Pogge}, {Stern}, {Treu}, \&
  {Woo}}]{Bentz2013}
{Bentz}, M.~C., {Denney}, K.~D., {Grier}, C.~J., {et~al.} 2013, \apj, 767, 149

\bibitem[{{Bentz} {et~al.}(2009){Bentz}, {Peterson}, {Netzer}, {Pogge}, \&
  {Vestergaard}}]{Ben2009}
{Bentz}, M.~C., {Peterson}, B.~M., {Netzer}, H., {Pogge}, R.~W., \&
  {Vestergaard}, M. 2009, \apj, 697, 160

\bibitem[{Brooks \& Gelman(1998)}]{Gel1998}
Brooks, S.~P. \& Gelman, A. 1998, Journal of Computational and Graphical
  Statistics, 7, 434

\bibitem[{{Burtscher} {et~al.}(2009){Burtscher}, {Jaffe}, {Raban},
  {Meisenheimer}, {Tristram}, \& {R{\"o}ttgering}}]{Bur2009}
{Burtscher}, L., {Jaffe}, W., {Raban}, D., {et~al.} 2009, \apjl, 705, L53

\bibitem[{{Crenshaw} {et~al.}(1999){Crenshaw}, {Kraemer}, {Boggess}, {Maran},
  {Mushotzky}, \& {Wu}}]{Cren1999}
{Crenshaw}, D.~M., {Kraemer}, S.~B., {Boggess}, A., {et~al.} 1999, \apj, 516,
  750

\bibitem[{{Czerny} {et~al.}(2003){Czerny}, {Doroshenko}, {Niko{\l}ajuk},
  {Schwarzenberg-Czerny}, {Loska}, \& {Madejski}}]{Czer2003}
{Czerny}, B., {Doroshenko}, V.~T., {Niko{\l}ajuk}, M., {et~al.} 2003, \mnras,
  342, 1222

\bibitem[{{Czerny} \& {Hryniewicz}(2011)}]{Czer2011}
{Czerny}, B. \& {Hryniewicz}, K. 2011, \aap, 525, L8

\bibitem[{{De Rosa} {et~al.}(2014)}]{DeRosa2014}
{De Rosa}, G., {et~al.} 2014, in preparation

\bibitem[{{Elitzur}(2006)}]{Eli2006}
{Elitzur}, M. 2006, \nar, 50, 728

\bibitem[{{Elvis} {et~al.}(2002){Elvis}, {Marengo}, \& {Karovska}}]{Elv2002}
{Elvis}, M., {Marengo}, M., \& {Karovska}, M. 2002, \apjl, 567, L107

\bibitem[{{Fanti} {et~al.}(1984){Fanti}, {Kellermann}, \& {Setti}}]{Fan1984}
{Fanti}, R., {Kellermann}, K., \& {Setti}, G. 1984, Science, 226, 473

\bibitem[{{Gaskell} {et~al.}(2004){Gaskell}, {Goosmann}, {Antonucci}, \&
  {Whysong}}]{Gask2004}
{Gaskell}, C.~M., {Goosmann}, R.~W., {Antonucci}, R.~R.~J., \& {Whysong}, D.~H.
  2004, \apj, 616, 147

\bibitem[{Gelman \& Rubin(1992)}]{Gel1992}
Gelman, A. \& Rubin, D.~B. 1992, Statistical science, 457

\bibitem[{{Grier} {et~al.}(2013){Grier}, {Peterson}, {Horne}, {Bentz}, {Pogge},
  {Denney}, {De Rosa}, {Martini}, {Kochanek}, {Zu}, {Shappee}, {Siverd},
  {Beatty}, {Sergeev}, {Kaspi}, {Araya Salvo}, {Bird}, {Bord}, {Borman}, {Che},
  {Chen}, {Cohen}, {Dietrich}, {Doroshenko}, {Efimov}, {Free}, {Ginsburg},
  {Henderson}, {King}, {Mogren}, {Molina}, {Mosquera}, {Nazarov}, {Okhmat},
  {Pejcha}, {Rafter}, {Shields}, {Skowron}, {Szczygiel}, {Valluri}, \& {van
  Saders}}]{Gri2013}
{Grier}, C.~J., {Peterson}, B.~M., {Horne}, K., {et~al.} 2013, \apj, 764, 47

\bibitem[{{Hawkins}(2002)}]{Hawk2002}
{Hawkins}, M.~R.~S. 2002, in Astronomical Society of the Pacific Conference
  Series, Vol. 284, IAU Colloq. 184: AGN Surveys, ed. R.~F. {Green}, E.~Y.
  {Khachikian}, \& D.~B. {Sanders}, 351

\bibitem[{{Heisler} {et~al.}(1997){Heisler}, {Lumsden}, \& {Bailey}}]{Hei1997}
{Heisler}, C.~A., {Lumsden}, S.~L., \& {Bailey}, J.~A. 1997, \nat, 385, 700
%

\bibitem[{{Hernitschek} {et~al.}(2014){Hernitschek}, {Rix}, {Bovy}, \&
  {Morganson}}]{Hern2014}
{Hernitschek}, N., {Rix}, H.-W., {Bovy}, J., \& {Morganson}, E. 2014, ArXiv
  e-prints

\bibitem[{{H{\"o}nig} \& {Kishimoto}(2011)}]{Hoe2011}
{H{\"o}nig}, S.~F. \& {Kishimoto}, M. 2011, \aap, 534, A121

\bibitem[{{Jaffe} {et~al.}(2004){Jaffe}, {Meisenheimer}, {R{\"o}ttgering},
  {Leinert}, {Richichi}, {Chesneau}, {Fraix-Burnet}, {Glazenborg-Kluttig},
  {Granato}, {Graser}, {Heijligers}, {K{\"o}hler}, {Malbet}, {Miley},
  {Paresce}, {Pel}, {Perrin}, {Przygodda}, {Schoeller}, {Sol}, {Waters},
  {Weigelt}, {Woillez}, \& {de Zeeuw}}]{Jaf2004}
{Jaffe}, W., {Meisenheimer}, K., {R{\"o}ttgering}, H.~J.~A., {et~al.} 2004,
  \nat, 429, 47

\bibitem[{{Kishimoto} {et~al.}(2011{\natexlab{a}}){Kishimoto}, {H{\"o}nig},
  {Antonucci}, {Barvainis}, {Kotani}, {Tristram}, {Weigelt}, \&
  {Levin}}]{Kish2011b}
{Kishimoto}, M., {H{\"o}nig}, S.~F., {Antonucci}, R., {et~al.}
  2011{\natexlab{a}}, \aap, 527, A121

\bibitem[{{Kishimoto} {et~al.}(2009){Kishimoto}, {H{\"o}nig}, {Antonucci},
  {Kotani}, {Barvainis}, {Tristram}, \& {Weigelt}}]{Kish2009}
{Kishimoto}, M., {H{\"o}nig}, S.~F., {Antonucci}, R., {et~al.} 2009, \aap, 507,
  L57

\bibitem[{{Kishimoto} {et~al.}(2013){Kishimoto}, {H{\"o}nig}, {Antonucci},
  {Millan-Gabet}, {Barvainis}, {Millour}, {Kotani}, {Tristram}, \&
  {Weigelt}}]{Kish2013}
{Kishimoto}, M., {H{\"o}nig}, S.~F., {Antonucci}, R., {et~al.} 2013, \apjl,
  775, L36

\bibitem[{{Kishimoto} {et~al.}(2011){Kishimoto}, {H{\"o}nig}, {Antonucci},
  {Millour}, {Tristram}, \& {Weigelt}}]{Kish2011a}
{Kishimoto}, M., {H{\"o}nig}, S.~F., {Antonucci}, R., {et~al.} 2011, \aap, 536,
  A78

\bibitem[{{Kishimoto} {et~al.}(2007){Kishimoto}, {H{\"o}nig}, {Beckert}, \&
  {Weigelt}}]{Kish2007}
{Kishimoto}, M., {H{\"o}nig}, S.~F., {Beckert}, T., \& {Weigelt}, G. 2007,
  \aap, 476, 713

\bibitem[{{Kollatschny} \& {Zetzl}(2013)}]{Kol2013}
{Kollatschny}, W. \& {Zetzl}, M. 2013, \aap, 551, L6

\bibitem[{{Koshida} {et~al.}(2009){Koshida}, {Yoshii}, {Kobayashi}, {Minezaki},
  {Sakata}, {Sugawara}, {Enya}, {Suganuma}, {Tomita}, {Aoki}, \&
  {Peterson}}]{Kosh2009}
{Koshida}, S., {Yoshii}, Y., {Kobayashi}, Y., {et~al.} 2009, \apjl, 700, L109

\bibitem[{{Kov{\'a}cs} {et~al.}(2004){Kov{\'a}cs}, {Mall}, {Bizenberger},
  {Baumeister}, \& {R{\"o}ser}}]{Kov2004}
{Kov{\'a}cs}, Z., {Mall}, U., {Bizenberger}, P., {Baumeister}, H., \&
  {R{\"o}ser}, H.-J. 2004, in Society of Photo-Optical Instrumentation
  Engineers (SPIE) Conference Series, Vol. 5499, Optical and Infrared Detectors
  for Astronomy, ed. J.~D. {Garnett} \& J.~W. {Beletic}, 432--441

\bibitem[{{Krolik} \& {Begelman}(1988)}]{Kro1988}
{Krolik}, J.~H. \& {Begelman}, M.~C. 1988, \apj, 329, 702

\bibitem[{{MacLeod} {et~al.}(2010){MacLeod}, {Ivezi{\'c}}, {Kochanek},
  {Koz{\l}owski}, {Kelly}, {Bullock}, {Kimball}, {Sesar}, {Westman}, {Brooks},
  {Gibson}, {Becker}, \& {de Vries}}]{MacL2010}
{MacLeod}, C.~L., {Ivezi{\'c}}, {\v Z}., {Kochanek}, C.~S., {et~al.} 2010,
  \apj, 721, 1014

\bibitem[{Mengersen \& Robert(2003)}]{Meng2003}
Mengersen, K. \& Robert, C. 2003, in {B}ayesian Statistics 7, ed. J.~Bernardo,
  M.~Bayarri, J.~Berger, A.~Dawid, D.~Heckerman, A.~Smith, \& M.~West (Oxford
  University Press, Oxford)

\bibitem[{{Minezaki} {et~al.}(2006){Minezaki}, {Yoshii}, {Aoki}, {Kobayashi},
  {Suganuma}, {Enya}, {Tomita}, \& {Peterson}}]{Sug2006}
{Minezaki}, T., {Yoshii}, Y., {Aoki}, T., {et~al.} 2006, in Astronomical
  Society of the Pacific Conference Series, Vol. 360, Astronomical Society of
  the Pacific Conference Series, ed. C.~M. {Gaskell}, I.~M. {McHardy}, B.~M.
  {Peterson}, \& S.~G. {Sergeev}, 79

\bibitem[{{Minezaki} {et~al.}(2004){Minezaki}, {Yoshii}, {Kobayashi}, {Enya},
  {Suganuma}, {Tomita}, {Aoki}, \& {Peterson}}]{Min2004}
{Minezaki}, T., {Yoshii}, Y., {Kobayashi}, Y., {et~al.} 2004, \apjl, 600, L35

\bibitem[{{Morganson} {et~al.}(2014){Morganson}, {Burgett}, {Chambers},
  {Green}, {Kaiser}, {Magnier}, {Marshall}, {Morgan}, {Price}, {Rix},
  {Schlafly}, {Tonry}, \& {Walter}}]{Morg2014}
{Morganson}, E., {Burgett}, W.~S., {Chambers}, K.~C., {et~al.} 2014, \apj, 784,
  92

\bibitem[{{Osterbrock} \& {Ferland}(2006)}]{Ost2006}
{Osterbrock}, D.~E. \& {Ferland}, G.~J. 2006, {Astrophysics of gaseous nebulae
  and active galactic nuclei}

\bibitem[{{Peng} {et~al.}(2002){Peng}, {Ho}, {Impey}, \& {Rix}}]{Pen2002}
{Peng}, C.~Y., {Ho}, L.~C., {Impey}, C.~D., \& {Rix}, H.-W. 2002, \aj, 124, 266

\bibitem[{{Peterson}(2006)}]{Pet2006a}
{Peterson}, B.~M. 2006, in Lecture Notes in Physics, Berlin Springer Verlag,
  Vol. 693, Physics of Active Galactic Nuclei at all Scales, ed. D.~{Alloin},
  77

\bibitem[{{Peterson} \& {Horne}(2006)}]{Pet2006b}
{Peterson}, B.~M. \& {Horne}, K. 2006, in Planets to Cosmology: Essential
  Science in the Final Years of the Hubble Space Telescope, ed. M.~{Livio} \&
  S.~{Casertano}, 89

\bibitem[{{Pott} {et~al.}(2010){Pott}, {Malkan}, {Elitzur}, {Ghez}, {Herbst},
  {Sch{\"o}del}, \& {Woillez}}]{Pot2010}
{Pott}, J.-U., {Malkan}, M.~A., {Elitzur}, M., {et~al.} 2010, \apj, 715, 736

\bibitem[{{Press} {et~al.}(1992){Press}, {Rybicki}, \& {Hewitt}}]{Pre1992}
{Press}, W.~H., {Rybicki}, G.~B., \& {Hewitt}, J.~N. 1992, \apj, 385, 404

\bibitem[{{Rayner} {et~al.}(2003){Rayner}, {Toomey}, {Onaka}, {Denault},
  {Stahlberger}, {Vacca}, {Cushing}, \& {Wang}}]{Ray2003}
{Rayner}, J.~T., {Toomey}, D.~W., {Onaka}, P.~M., {et~al.} 2003, \pasp, 115,
  362

\bibitem[{{Riffel} {et~al.}(2009){Riffel}, {Storchi-Bergmann}, \&
  {McGregor}}]{Rif2009}
{Riffel}, R.~A., {Storchi-Bergmann}, T., \& {McGregor}, P.~J. 2009, \apj, 698,
  1767

\bibitem[{{Rybicki} \& {Press}(1992)}]{Ryb1992}
{Rybicki}, G.~B. \& {Press}, W.~H. 1992, \apj, 398, 169

\bibitem[{{Schartmann} {et~al.}(2010){Schartmann}, {Burkert}, {Krause},
  {Camenzind}, {Meisenheimer}, \& {Davies}}]{Scha2010}
{Schartmann}, M., {Burkert}, A., {Krause}, M., {et~al.} 2010, \mnras, 403, 1801

\bibitem[{{Schmidt} {et~al.}(2010){Schmidt}, {Marshall}, {Rix}, {Jester},
  {Hennawi}, \& {Dobler}}]{Schm2010}
{Schmidt}, K.~B., {Marshall}, P.~J., {Rix}, H.-W., {et~al.} 2010, \apj, 714,
  1194

\bibitem[{{Schn{\"u}lle} {et~al.}(2013){Schn{\"u}lle}, {Pott}, {Rix},
  {Decarli}, {Peterson}, \& {Vacca}}]{Schn2013}
{Schn{\"u}lle}, K., {Pott}, J.-U., {Rix}, H.-W., {et~al.} 2013, \aap, 557, L13

\bibitem[{{Shappee} {et~al.}(2014){Shappee}, {Prieto}, {Grupe}, {Kochanek},
  {Stanek}, {De Rosa}, {Mathur}, {Zu}, {Peterson}, {Pogge}, {Komossa}, {Im},
  {Jencson}, {Holoien}, {Basu}, {Beacom}, {Szczygie{\l}}, {Brimacombe},
  {Adams}, {Campillay}, {Choi}, {Contreras}, {Dietrich}, {Dubberley},
  {Elphick}, {Foale}, {Giustini}, {Gonzalez}, {Hawkins}, {Howell}, {Hsiao},
  {Koss}, {Leighly}, {Morrell}, {Mudd}, {Mullins}, {Nugent}, {Parrent},
  {Phillips}, {Pojmanski}, {Rosing}, {Ross}, {Sand}, {Terndrup}, {Valenti},
  {Walker}, \& {Yoon}}]{Shap2014}
{Shappee}, B.~J., {Prieto}, J.~L., {Grupe}, D., {et~al.} 2014, \apj, 788, 48

\bibitem[{{Shappee} \& {Stanek}(2011)}]{Shappee2011}
{Shappee}, B.~J. \& {Stanek}, K.~Z. 2011, \apj, 733, 124

\bibitem[{{Siverd} {et~al.}(2012){Siverd}, {Beatty}, {Pepper}, {Eastman},
  {Collins}, {Bieryla}, {Latham}, {Buchhave}, {Jensen}, {Crepp}, {Street},
  {Stassun}, {Gaudi}, {Berlind}, {Calkins}, {DePoy}, {Esquerdo}, {Fulton},
  {F{\H u}r{\'e}sz}, {Geary}, {Gould}, {Hebb}, {Kielkopf}, {Marshall}, {Pogge},
  {Stanek}, {Stefanik}, {Szentgyorgyi}, {Trueblood}, {Trueblood}, {Stutz}, \&
  {van Saders}}]{Siverd2012}
{Siverd}, R.~J., {Beatty}, T.~G., {Pepper}, J., {et~al.} 2012, \apj, 761, 123

\bibitem[{{Skrutskie} {et~al.}(2006){Skrutskie}, {Cutri}, {Stiening},
  {Weinberg}, {Schneider}, {Carpenter}, {Beichman}, {Capps}, {Chester},
  {Elias}, {Huchra}, {Liebert}, {Lonsdale}, {Monet}, {Price}, {Seitzer},
  {Jarrett}, {Kirkpatrick}, {Gizis}, {Howard}, {Evans}, {Fowler}, {Fullmer},
  {Hurt}, {Light}, {Kopan}, {Marsh}, {McCallon}, {Tam}, {Van Dyk}, \&
  {Wheelock}}]{Skru2006}
{Skrutskie}, M.~F., {Cutri}, R.~M., {Stiening}, R., {et~al.} 2006, \aj, 131,
  1163

\bibitem[{{Ter Braak}(2006)}]{Braak2006}
{Ter Braak}, C.~J.~F. 2006, Statistics and Computing, 16, 239

\bibitem[{{Tr{\`e}vese} {et~al.}(2001){Tr{\`e}vese}, {Kron}, \&
  {Bunone}}]{Trev2001}
{Tr{\`e}vese}, D., {Kron}, R.~G., \& {Bunone}, A. 2001, \apj, 551, 103

\bibitem[{{Urry} \& {Padovani}(1995)}]{Urr1995}
{Urry}, C.~M. \& {Padovani}, P. 1995, \pasp, 107, 803

\bibitem[{{Zu} {et~al.}(2011){Zu}, {Kochanek}, \& {Peterson}}]{Zu2011}
{Zu}, Y., {Kochanek}, C.~S., \& {Peterson}, B.~M. 2011, \apj, 735, 80

\end{thebibliography}

\begin{appendix}
\section{Gelman convergence diagnostics}
\label{sec:Gelman}

We check for convergence using the Gelman convergence diagnostics 
\citep{Gel1992, Gel1998}.
Given $m$ independent\footnote{The DE-MC proposal scheme presented in Eq.~\ref{eq:DEMC} might at first glance seem to violate one of the basic assumptions for monitoring convergence with the $\hat{R}$-statistic of \citet{Gel1992, Gel1998}, namely the assumtion that the $m$ individual chains are independent of each other. However, \citet{Braak2006} demonstrates that the conditional stationary pdf of one individual chain does not depend on the states of the other chains and is identical for all chains, so that the joint stationary pdf $p(\mathbf{x}_1,...,\mathbf{x}_m)$ of the $m$ chains is given simply by the product $p(\mathbf{x}_1)\times ... \times p(\mathbf{x}_m)$. Thus, the states of the individual chains $\mathbf{x}_1,...,\mathbf{x}_m$ are independent of each other at any iteration step after the burn-in phase \citep{Braak2006, Meng2003}, i.e., after the algorithm has become independent of the starting distribution. Convergence of the DE-MC algorithm can therefore be monitored using Gelman's convergence criterion.} chains and $2n$ iterations of the sampler, we can calculate for the second half of the iterations\footnote{Only the second half is evaluated in order to lower effects of the influence of the starting distribution, i.e., the burn-in phase is discarded.} in each step and for each parameter estimand $x$ the potential scale reduction factor (PSRF)
\begin{equation}
\hat R=\frac{\hat{V}}{W}=\frac{\hat \sigma_+^2+B/(mn)}{W}\, ,
\label{eq:Rhat}
\end{equation}
which serves as an (over-) estimate of the true scale reduction factor $R=\hat{V}/\sigma^2$. Here, 
\begin{equation}
\hat \sigma_+^2=\frac{n-1}{n}W+\frac{B}{n}
\label{eq:sigma}
\end{equation}
is an estimator of the true target variance $\sigma^2$ and is calculated by a weighted
mean of the between-chain variance $B/n$
\begin{equation}
B/n=\frac{1}{m-1}\sum_{j=1}^{m}(\bar{x}_{j.}-\bar{x}_{..})^2 \, ;
\end{equation}
i.e.~the variance between the $m$ chain means $\bar{x}_{j.}$ (where $\bar{x}_{..}$ denotes the global mean over all chains and iterations),
and the pooled within-chain variance $W$:
\begin{equation}
W=\frac{1}{m(n-1)}\sum_{j=1}^{m}\sum_{t=1}^{n}(\bar{x}_{jt}-\bar{x}_{j.})^2\, .
\label{eq:W}
\end{equation}
The true target mean $\mu$ is estimated by the global sample mean of $x$, i.e.~$\hat \mu=\bar x_{..}$, and the term $B/(mn)$ in Eq.~\ref{eq:Rhat} refers to the sampling 
variability of $\hat \mu$.
As $W$ always underestimates the true variance ($W \to \sigma^2$ for $n \to \infty$) for any finite $n$, the PSRF $\hat R$ will always overestimate the true scale reduction factor and can be
used as convergence diagnostic:\\
High values of $\hat R$, i.e.~values significantly above 1, indicate that further simulations may lead to an improved inference of the target distribution -- either because the variance estimates in the numerator of Eq.~\ref{eq:Rhat} can be decreased further by running more iterations or because $W$ will increase with continuing iterations, because the chains have not yet covered the complete target distribution. If $\hat R$ is close to 1, it can be assumed that $W$ has nearly converged to the target variance and that each of the $m$ chains of $n$ iterations is sufficiently close to the target distribution.\\
The above holds if the sampler is started with a sufficiently overdispersed starting distribution. Otherwise $\hat R$ values close to 1 might falsely diagnose convergence \citep{Gel1998}. Therefore, it is strongly recommended to graphically inspect the evolution of $W$ and $\hat V$, to make certain that they are not still evolving.

\end{appendix}
% ...

\end{document}